\documentclass[conference]{IEEEtran}
\IEEEoverridecommandlockouts
\usepackage{fancyhdr}
\usepackage{cite}
\usepackage{amsmath,amssymb,amsfonts}
\usepackage{algorithmic}
\usepackage{graphicx}
\usepackage{textcomp}
\usepackage{xcolor}
\usepackage{array}    
\usepackage{fancyhdr}
\usepackage{graphicx}
\usepackage{amsmath}
\usepackage{array}
\usepackage{amsfonts}
\usepackage{multirow}
\usepackage{booktabs}

\def\BibTeX{{\rm B\kern-.05em{\sc i\kern-.025em b}\kern-.08em
    T\kern-.1667em\lower.7ex\hbox{E}\kern-.125emX}}

\usepackage{titlesec}

\usepackage{amsmath}

\renewcommand{\thesection}{\arabic{section}}
\renewcommand{\thesubsection}{\arabic{section}.\arabic{subsection}}
\renewcommand{\thesubsubsection}{\arabic{section}.\arabic{subsection}.\arabic{subsubsection}}

\titleformat{\section}[block]{\bfseries\normalsize}{\thesection\quad}{0pt}{}  
\titleformat{\subsection}[block]{\itshape\normalsize}{\thesubsection\quad}{0pt}{}  
\titleformat{\subsubsection}[block]{\itshape\normalsize}{\thesubsubsection\quad}{0pt}{} 

\begin{document}
\rmfamily

\title{{\fontsize{16}{19}\selectfont \textbf{Evaluation of the Impact of IBR on the Frequency Dynamics in the Brazilian Power System}}}

\author{

    \IEEEauthorblockN{{\fontsize{12}{14}\selectfont Bruno Pinheiro}}
    \IEEEauthorblockA{
    \textit{\fontsize{10}{12}\selectfont Universidade Estadual de Campinas}\\
    \fontsize{10}{12}\selectfont Campinas, Brazil \\
    \fontsize{10}{12}\selectfont b229989@dac.unicamp.br}
    \and
   
    \IEEEauthorblockN{{\fontsize{12}{14}\selectfont M. C. Llerena Velasquez}}
    \IEEEauthorblockA{
    \textit{\fontsize{10}{12}\selectfont Universidade Estadual de Campinas}\\
    \fontsize{10}{12}\selectfont Campinas, Brazil \\
    \fontsize{10}{12}\selectfont m272292@dac.unicamp.br}
    \and
   
    \IEEEauthorblockN{{\fontsize{12}{14}\selectfont Giovane Faria}}
    \IEEEauthorblockA{
    \textit{\fontsize{10}{12}\selectfont Federal University of Santa Catarina}\\
    \fontsize{10}{12}\selectfont Florianópolis, Brazil \\
    \fontsize{10}{12}\selectfont gio96ifsc@gmail.com}
    \and
    
    \IEEEauthorblockN{{\fontsize{12}{14}\selectfont A. F. C. Aquino}}
    \hspace{6.5cm}\IEEEauthorblockA{
    \textit{\fontsize{10}{12}\selectfont Federal University of Santa Catarina}\\
    \fontsize{10}{12}\selectfont Florianópolis, Brazil \\
    \textit{\fontsize{10}{12}\selectfont INESC P\&D Brasil}\\
    \fontsize{10}{12}\selectfont Santos, Brazil \\
    \fontsize{10}{12}\selectfont antoniofelipeaquino25@gmail.com}    
    \and
    
    \IEEEauthorblockN{{\fontsize{12}{14}\selectfont Diego Issicaba}}
    \IEEEauthorblockA{
    \textit{\fontsize{10}{12}\selectfont Federal University of Santa Catarina}\\
    \fontsize{10}{12}\selectfont Florianópolis, Brazil \\
    \textit{\fontsize{10}{12}\selectfont INESC P\&D Brasil}\\
    \fontsize{10}{12}\selectfont Santos, Brazil \\
    \fontsize{10}{12}\selectfont diego.issicaba@inescbrasil.org.br}    
    \and
    
    \IEEEauthorblockN{{\fontsize{12}{14}\selectfont Daniel Dotta}}
    \IEEEauthorblockA{
    \textit{\fontsize{10}{12}\selectfont Universidade Estadual de Campinas}\\
    \fontsize{10}{12}\selectfont Campinas, Brazil \\
    \fontsize{10}{12}\selectfont dottad@unicamp.br}
}
\maketitle
\thispagestyle{fancy}

\begin{abstract}
In this paper, the impact of inverter-based resources (IBRs) on the frequency dynamics of the Brazilian Interconnected Power System (BIPS) is evaluated. A measurement-based framework is proposed to assess the impact of IBR penetration on the system-wide and regional/local frequency dynamic. The analysis leverages data from a low-voltage wide area monitoring system (WAMS) and publicly available historical generation records from the Brazilian System Operator. A methodology is introduced to extract local frequency fluctuations across regions using a variational mode decomposition (VMD) approach. The findings reveal a continuous degradation in the system-wide frequency and local frequency variations, underscoring the need for enhanced regional monitoring and evaluation metrics to maintain frequency stability in large-scale interconnected systems.
\end{abstract}


\section{Introduction}

\subsection{Motivation}

With the increasing integration of inverter-based resources (IBRs) and the gradual replacement of conventional generation, significant challenges emerge in the planning, operation, and control of power systems. The analysis of dynamic frequency behavior has become crucial due to the low inertia characteristics and the stochastic nature of IBRs, such as wind and solar generation. Frequency stability issues are especially prevalent in smaller radial systems with high IBR penetration, like those in Great Britain, Ireland, and Australia. However, on August 15, 2023, the Brazilian Interconnected Power System (BIPS)—a large interconnected network—experienced a blackout triggered by a 500 kV transmission line trip in the Northeast region, which subsequently led to the automatic tripping of interconnections between the Northeast, North,
and Southeast regions, due to the operation of out-of-step protections~\cite{RAP}. While there is no conclusive evidence linking this event to frequency stability, it is widely understood that the high level of IBR penetration at the time of the disturbance played a significant role. This paper aims to investigate the impact of IBR penetration on the dynamic frequency behavior of large interconnected systems like BIPS.

\subsection{Literature Review}

The degradation of frequency quality has been reported in power systems worldwide. For example, due to the increasing penetration of IBR, the Australian Energy Market Operator (AEMO) documented a progressive decline in frequency response in both mainland and Tasmania systems under normal operating conditions~\cite{AEMC2020}. Moreover, specific areas within the system have shown more pronounced frequency fluctuations~\cite{Australia_TPWRS}. In the British power system, the frequency variance increased by approximately 0.015 Hz from 2014 to 2019~\cite{Cai2024}. Additionally, the power systems in Ireland and Northern Ireland exhibited a linear increase in the frequency standard deviation over the last three years (2020-2023)~\cite{Kerci2023}.

The correlation between the penetration of IBR and the rate of change of frequency (RoCoF) and frequency deviation in Australia and Germany is investigated in~\cite{Xypolytou2018}. Similarly,~\cite{Adeen2019} presents the correlation between the high penetration of wind generation and the frequency behavior of the Ireland system. However, most of these studies focus on frequency variations of the center of inertia (COI), overlooking the spatio-temporal characteristics of frequency dynamics~\cite{Milano_FDF}. Notably, the systems evaluated in these studies lack the continental scale and interconnected complexity of the BIPS, underscoring the importance of assessing the impacts of IBR at a local level.

Previous research has examined the influence of the stochastic behavior of IBR, loads, and controls on the frequency distribution by statistical analyses of frequency measurements~\cite{Giudice2021, Francesca2016, Zuo2021, Vorobev2019}. Recently, analytical methods have been proposed to decompose frequency into its local and global dynamic components~\cite{Bruno2023,Jiaxin2024,Licheng2024}. In the literature, methodologies based on time-domain simulation have been investigating the impact of IBR integration on frequency~\cite{Guilherme2024} and voltage stability~\cite{David2024} in the BIPS. In this paper, real-world measurements are used to evaluate the IBR impact on both the system-wide and local frequency dynamics of the BIPS. To isolate local frequency fluctuations, we apply the variational mode decomposition (VMD) method to local frequency measurements from the BIPS. Based on these measurements, we provide a comprehensive assessment of the influence of IBR penetration on both system-wide and local frequency variations.

\subsection{Contributions}

The contributions of this work can be summarized as follows:
\begin{itemize} 
    \item A measurement-based framework that utilizes real-world PMU data to evaluate both system-wide and local frequency dynamic behavior in a large-scale power system, specifically the BIPS;

    \item Clear indication of an increasing trend in system-wide overfrequency events in the BIPS over the past decade, primarily driven by overgeneration from IBR during periods of high IBR penetration.  

    \item Evidence of increased frequency volatility at both the system-wide and local levels, with a pronounced impact in regions characterized by lower regional inertia.  
    \end{itemize}

\subsection{Organization}

The structure of this paper is as follows. Section 2 provides background information on system-wide and local dynamic frequency behavior, along with the proposed method for assessing local frequency variations. Section 3 introduces the BIPS and details the metrics of IBR penetration observed in 2023. Section 4 presents an evaluation of the system-wide frequency dynamics from 2011 to 2023, while Section 5 focuses on the local frequency analysis for 2023. Section 6 discusses the observed degradation of frequency quality in the BIPS. Finally, Section 7 concludes with a summary and closing remarks.

\section{System-wide and Local Frequency}

This section introduces the problem of spatio-temporal frequency dynamics, distinguishing between system-wide and local frequency in power systems. It also presents the variational mode decomposition approach used to assess local frequency fluctuations.

\subsection{System-wide and Local frequency}

In power systems dominated by synchronous generators, the system-wide dynamic frequency behavior is represented by the center of inertia (COI) frequency, defined as follows:
\begin{equation}
    \label{eq:coi}
    \omega_{COI} = \frac{\sum_{k=1}^{n_g}H_k\omega_k}{\sum_{k=1}^{n_g}H_k}
\end{equation}

\noindent where $\omega_k$ and $H_k$ are the internal frequency and inertia constant of machine $k$, respectively. The COI frequency captures the average frequency behavior across the system. However, frequency dynamics are known to exhibit significant spatio-temporal characteristics~\cite{Milano_FDF}.

In low-inertia systems, local frequency variations tend to deviate more from the COI frequency due to reduced inertia levels and the uneven distribution of inertia across the system. Regions with low inertia are more susceptible to significant frequency fluctuations and high local RoCoFs following disturbances. Even under normal operating conditions, local frequency variations can be more pronounced compared to other regions, potentially indicating weak areas in terms of frequency control capability. Additionally, real-world measurements show that power systems are continuously subjected to load variations and stochastic inputs from IBRs, such as wind and solar generation, which contribute to locational variability in frequency dynamics. Moreover, local frequency behavior is influenced by regional electromechanical oscillatory modes~\cite{Guido2022} and rapid oscillations induced by IBRs~\cite{lingling2023}.

In a large interconnected power system, the spatio-temporal characteristics of frequency become even more critical.  Regional weaknesses can impact the entire system, as local disturbances may propagate across the network, potentially leading to regional separation and blackouts. Evaluating regional frequency strength in a large-scale power system presents significant challenges. Therefore, analyzing the statistical properties of local frequency dynamics is essential for identifying regions with lower frequency strength. These insights can support targeted actions to mitigate potential local issues and enhance overall system resilience.

\subsection{Extraction of Local Frequency Fluctuations}

The frequency dynamics recorded by a phasor measurement unit (PMU) during ambient conditions or after a disturbance can be decomposed into two components~\cite{Jiaxin2024,Osipov2023}: (i) the COI frequency, or quasi-steady-state (QSS), and (ii) local frequency fluctuations, which include dynamic components such as oscillations and stochastic variations. Hereafter, we refer to the system-wide frequency as the COI/QSS component and to local frequency fluctuations as the dynamic component of the local frequency.

To effectively extract local frequency fluctuations from PMU measurements, we propose the application of a variational mode decomposition (VMD) approach. VMD decomposes a signal into a set of intrinsic mode functions (IMFs), or modes~\cite{Mario2019, Zamora2019}:
\begin{equation}
    \label{eq:modes}
    s(t) = \sum_{k=1}^{n} s_k(t) + r_n(t)
\end{equation}
\noindent where $s(t)$ is the original signal, $s_k(t)$ represents the $k$-th mode, $n$ is the number of modes, and $r_n(t)$ is the residue or noise. When the IMFs are arranged from the slowest to the fastest varying components, the first IMF corresponds to the quasi-steady-state (QSS) component of the signal. The VMD estimates the predefined number of modes, $n$, through a constrained variational optimization problem, as follows:

\begin{equation}
\label{eq:vmd}
\begin{aligned}
\min_{\{s_k\},\{\omega_k\}} &\left\{ \sum_{k=1}^{n} \left\| \frac{\partial}{\partial t} \left[ \left( \delta(t) + \frac{j}{\pi t} \right) * s_k(t) \right] e^{-j\omega_k t} \right\|_2^2 \right\},  \\
\text{s.t.} &\sum_{k=1}^{n} s_k(t) = \mathbf{s}(t).
\end{aligned}
\end{equation}

\noindent where $s_k$ and $\omega_k$ are the k-\textit{th} mode and its center frequency, respectively; $\delta$ is the Dirac distribution, and $*$ denotes convolution operation.

The optimization problem in \eqref{eq:vmd} is solved using the alternating direction method of multipliers (ADMM). To address the constrained variational optimization problem, several key parameters must be defined: the number of modes ($n$), the tolerance factor for convergence ($\epsilon$), the initial center frequencies ($\mathbf{\omega}_k$), and the balancing parameter of the data-fidelity constraint ($\alpha$). For further details on the mathematical formulation and algorithm for solution of VMD, readers are referred to~\cite{Dragomiretskiy2014, Mario2019}.

Although the VMD approach requires the selection of hyperparameters, it offers several advantages over similar methods. Notably, VMD demonstrates superior performance in estimating the quasi-steady-state (QSS) component by capturing not only trends but also aperiodic variations~\cite{Osipov2023}. For frequency measurements collected by PMUs located at various points in the system, which are subject to system load variations and aperiodic variations induced by IBR dynamics, VMD enables improved extraction of dynamic components~\cite{Osipov2023}. Additionally, other decomposition techniques, such as Empirical Mode Decomposition (EMD), are prone to mode mixing due to their empirical nature. In contrast, VMD formulates the decomposition as an optimization problem in the spectral domain, ensuring better mode separation and greater stability in decomposition results. In this study, a Python-based toolbox is used to extract intrinsic mode functions (IMFs) through VMD~\cite{Carvalho2020}.

\begin{figure}[htb]
{\centering
\includegraphics[width=1\linewidth]{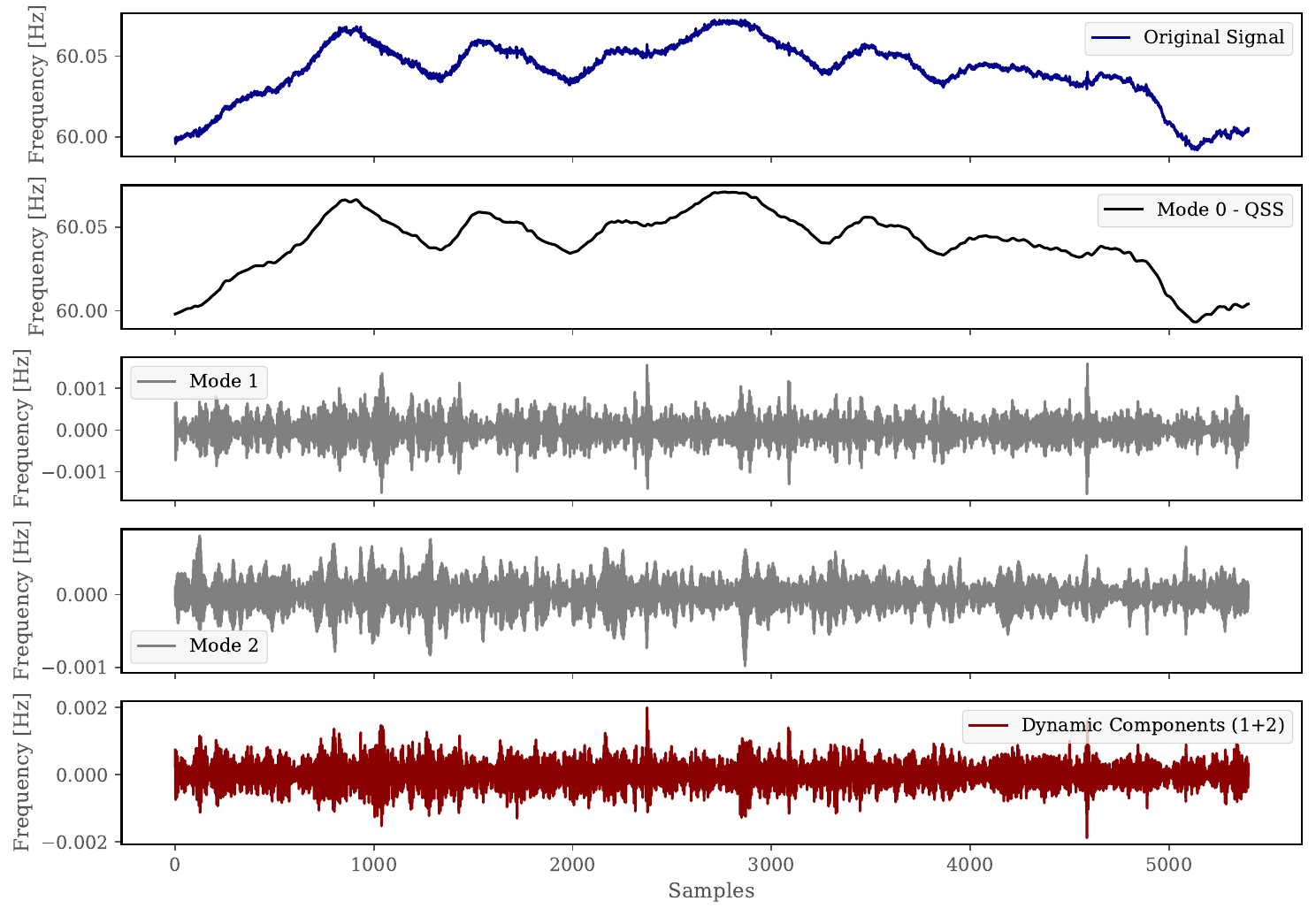}
\caption{Example of application of VMD on ambient frequency data to extract the QSS and dynamic component using three modes.}
\label{fig:exampleVMD}}
\end{figure}

As an example, Figure~\ref{fig:exampleVMD} illustrates the application of VMD on an ambient frequency signal measured in the BIPS. Mode 0 represents the QSS component, reflecting the system-wide frequency behavior primarily influenced by system load variations. Modes 1 and 2 capture the dynamic components, which are primarily associated with stochastic load variations and IBR activity near the measurement point. The signal in red color is the dynamic component of the frequency signal, that is, the sought local frequency fluctuation. Based on empirical testing with both ambient data and disturbance events, the number of modes was determined to be three. The application of the VMD approach for local frequency assessment is further detailed in Section V.

\section{Brazilian Interconnected Power System (BIPS)}
\label{sec:BIPS}

This section provides an overview of the BIPS, including its generation matrix, geo-electrical regions, and monitoring infrastructure.

\subsection{Generation Profile and Regional Characteristics of the BIPS}

\begin{figure}[t]
{\centering
\includegraphics[width=1\linewidth]{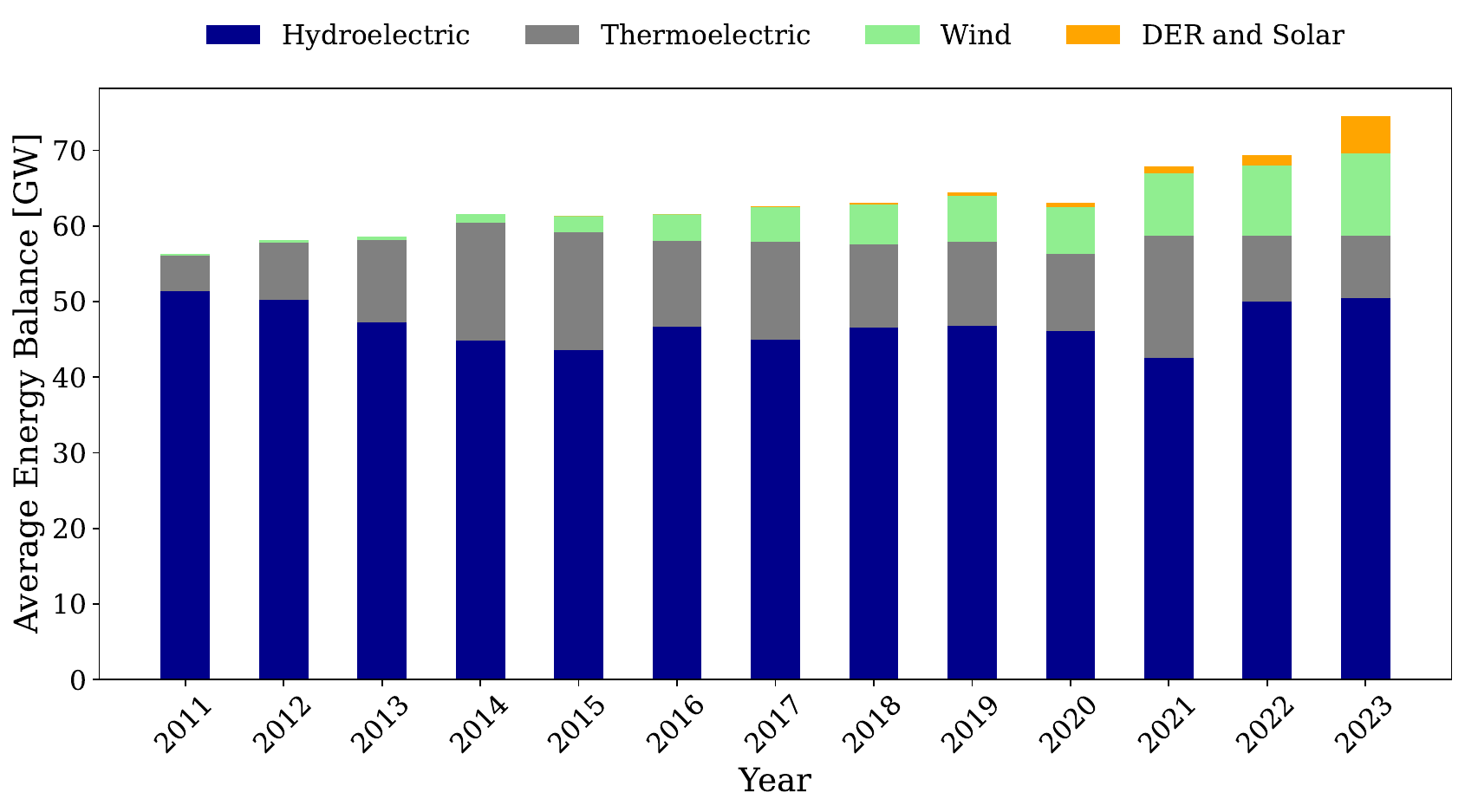}
\caption{Average annual generation by generation type over the years in the BIPS.}
\label{fig:capacity}}
\end{figure}

The BIPS is a bulk and meshed power system with continental dimension. The primary generation of the BIPS is hydroelectric, comprising most of its composition. In 2023, the total installed capacity of the BIPS is 212.7 GW, composed of 51.0\% hydro, 12.9\% wind, 11.3\% thermal, 11.3\% distributed generation (DER), 7.3\% biomass, 5.1\% solar, and 0.9\% nuclear~\cite{ONS_CAPACITY}. The integration of wind and solar energy into the BIPS has grown significantly in recent years, as illustrated in the average annual energy balance in Figure~\ref{fig:capacity}. Notably, the wind generation has experienced exponential participation in the BIPS since 2014, reaching approximately 10.9 GW (14.6\%) by 2023. Similarly, solar generation began increasing its share around 2017, surpassing 4.9 GW (6.6\%) in 2023.

\begin{figure}[t]
{\centering
\includegraphics[width=0.75\linewidth]{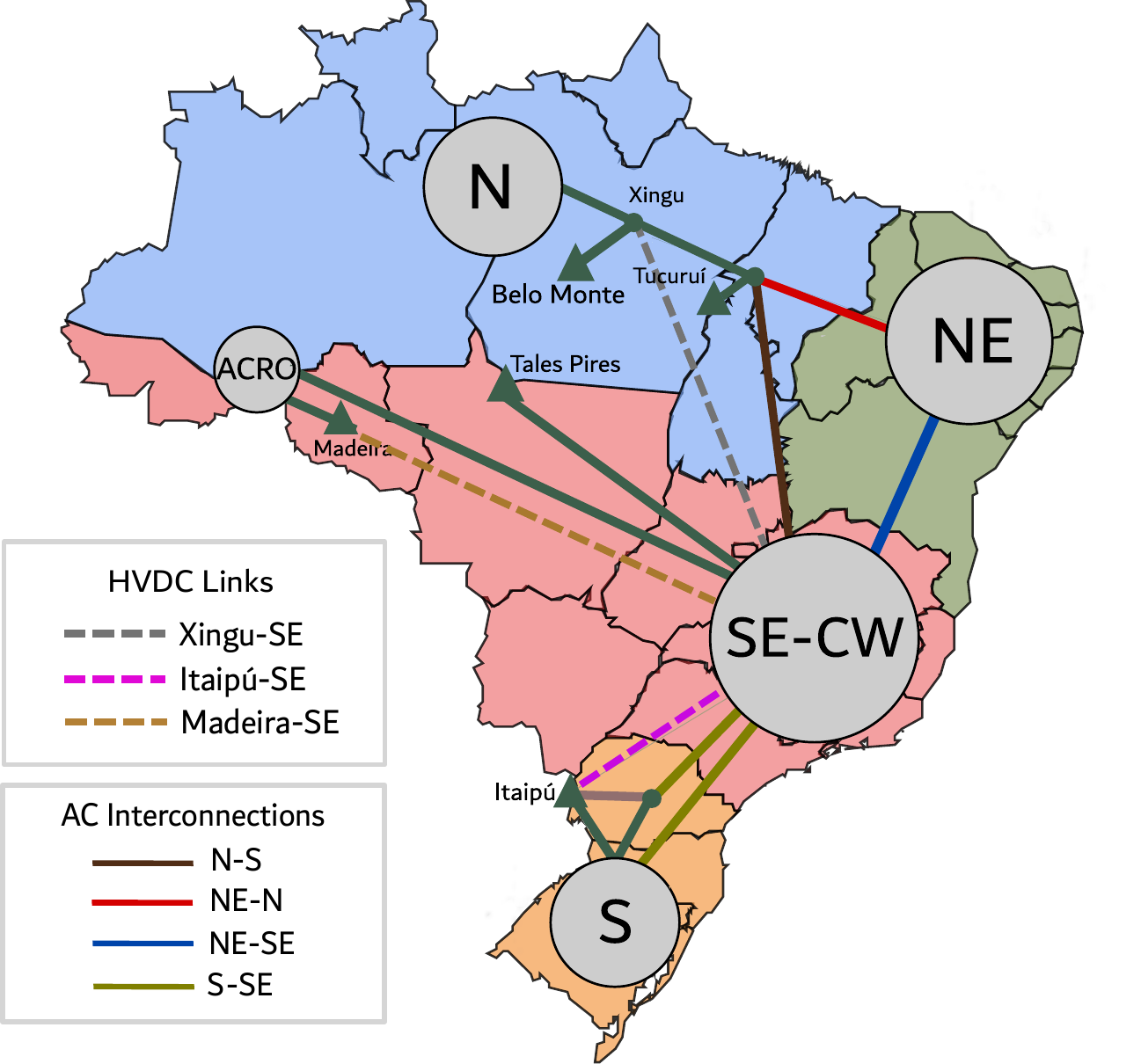}
\caption{BIPS geo-electrical regions and main interconnections}
\label{fig:mapa}}
\end{figure}

The BIPS is divided into four geo-electric regions: North (N), Northeast (NE), Southeast-Central West (SE-CW), and South (S). The geo-electric regions within Brazil and its interconnections are illustrated in Figure~\ref{fig:mapa}, including the six HVDC bipoles interconnecting the SE-CW with North, and South areas. The average annual generation and demand for each region in 2023 are summarized in Table~\ref{tab:average_regions}. The SE-CW region serves as the primary load center of the BIPS, with a demand exceeding the combined demand of the other regions and significant hydroelectric generation. Both SE–CW and NE regions exhibit a high concentration of solar generation, encompassing both centralized and distributed systems. Additionally, the NE region has experienced remarkable growth in wind generation, reaching an installed capacity of over 23 GW in 2023.

\begin{table}[ht]
\centering
\caption{Average annual generation and demand by geo-electric region in 2023 [GW].}
\label{tab:average_regions}
\begin{tabular}{cccccc}
\hline
\textbf{Region} & \textbf{Hydraulic} & \textbf{Thermal} & \textbf{Wind} & \textbf{Solar*} & \textbf{Demand} \\ \hline
N & 7.969 & 1.420 & 0.210 & 0.208 & 7.140 \\ 
NE & 4.018 & 0.540 & 10.009 & 1.821 & 12.117 \\ 
S & 9.288 & 1.034 & 0.662 & 0.561 & 12.567 \\ 
SE-CW & 29.192 & 5.256 & 0.007 & 2.319 & 41.881 \\ 
\hline
\end{tabular}
\label{tab:avg_gen_2023}
\begin{flushleft}
\footnotesize{\textit{*The solar generation values include both centralized and distributed solar generation (DER).}}
\end{flushleft}
\end{table}

\subsection{PMU measurements in BIPS}

The BIPS is monitored through PMUs installed at the high-voltage level under the supervision of the Brazilian System Operator (SO). Additionally, an independent PMU system known as the MedFasee project, led by the Federal University of Santa Catarina (UFSC) in Brazil~\cite{DeckerPMU}~\cite{Vaz2021}, encompasses 28 PMUs strategically placed at universities and connected to public outlets at low-voltage level. This system is dedicated to monitoring voltage synchrophasors and frequency, with a rate of 60 phasors per second. In this study, the frequency dataset from the MedFasee project will be used. 

\subsection{IBR penetration in BIPS (2023)}

In this paper, we are interested in investigating the impact of IBR penetration in the system-wide frequency response and also in the local frequency variations. Thus, the IBR system-wide penetration and regional penetration are calculated. Public data on regional demand and generation disclosed by the Brazilian SO have been used~\cite{OperacaoSIN}. The IBR penetration is calculated as follows:
\begin{equation}
\label{eq:IBR_regional}
\text{IBR}^i_h\% = \frac{P_{\text{wind}}^{i} + P_{\text{solar}}^{i} + P_{\text{DER}}^{i}}{P_{L}^{i}} \times 100
\end{equation}
\noindent where, $\text{IBR}^i_h\%$ is the regional penetration of the geoelectric area $i$, during hour $h$, with $P_{\text{wind}}^{i}$ representing the wind active power, $P_{\text{solar}}^{i}$ denoting solar active power, $P_{\text{DER}}^{i}$ referring to distributed energy resource (DER) active power, and $P_{L}^{i}$ representing the demand of the region $i$.

\begin{figure}[t]
{\centering
\includegraphics[width=1\linewidth]{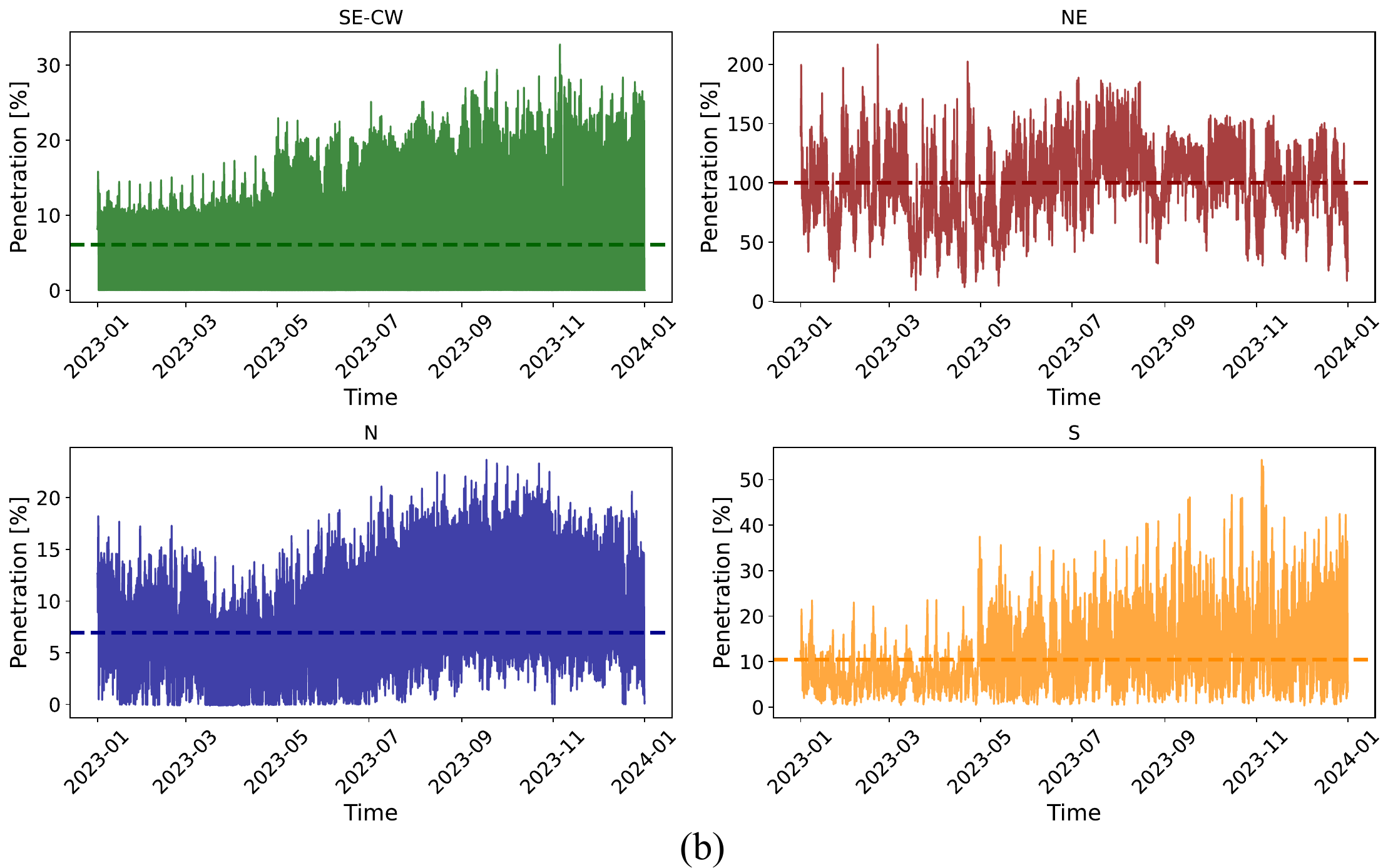}
\caption{Regional IBR penetration in the BIPS in 2023 with hourly resolution.}
\label{fig:regional_ibr}}
\end{figure}

The average level of IBR penetration in the BIPS for 2023 is around 25\%, peaking at around 48\% between June and August. The average hourly IBR penetration by region is shown in Figure~\ref{fig:regional_ibr}, with the Northeast region exhibiting the highest penetration relative to local demand, followed by the South region. There is a notable disparity between these regions, with the Northeast reaching penetration of 150\% IBR on some days. This indicates that the Northeast can not only meet its local demand, but also export surplus generation using IBRs. For instance, on August 15, 2023 (BIPS Blackout), the Northeast recorded about 198\% regional IBR penetration~\cite{RAP}. The maximum average daily IBR penetration and the annual average for each are detailed in Table~\ref{tab:average_hour_BIPS}.

\begin{table}[htb]
    \centering
    \caption{Max. Average daily IBR penetration by region in BIPS in 2023}
    \label{tab:average_hour_BIPS}
    \begin{tabular}{lcccc}
        \toprule
        \multirow{2}{*}{\textbf{Region}} & \multicolumn{2}{c}{\textbf{Maximum Penetration} [\%]} & \multirow{2}{*}{\textbf{Avg.} \textbf{Penetration} [\%]} \\
        \cmidrule(lr){2-3}
        & \textbf{Day} & \textbf{Penetration} \\
        \midrule
        NE & 2023-07-04 & 216.8 & 100.1 \\
        SE \& CW & 2023-11-11 & 32.7 & 6.1 \\
        N & 2023-10-01 & 23.7 & 6.9 \\
        S & 2023-06-16 & 54.4 & 10.5 \\
        \bottomrule
    \end{tabular}
\end{table}

The Brazilian SO has been monitoring synchronous inertia through state estimation, which involves tracking the status of generation units—whether they are online or offline. It is important to note that in the BIPS, wind and solar generation operate in grid-following mode. The BIPS grid code specifies that wind generators must provide during underfrequency events a frequency response equivalent to 10\% of their nominal active power when the measured frequency deviates by 0.2 Hz from the nominal 60 Hz. This response is referred to in the grid code as \textit{synthetic inertia}. However, since there is no inertial contribution during normal operating conditions or within the defined frequency threshold, this response is not considered in the estimating of inertia levels.

\begin{figure}[htb]
{\centering
\includegraphics[width=1\linewidth]{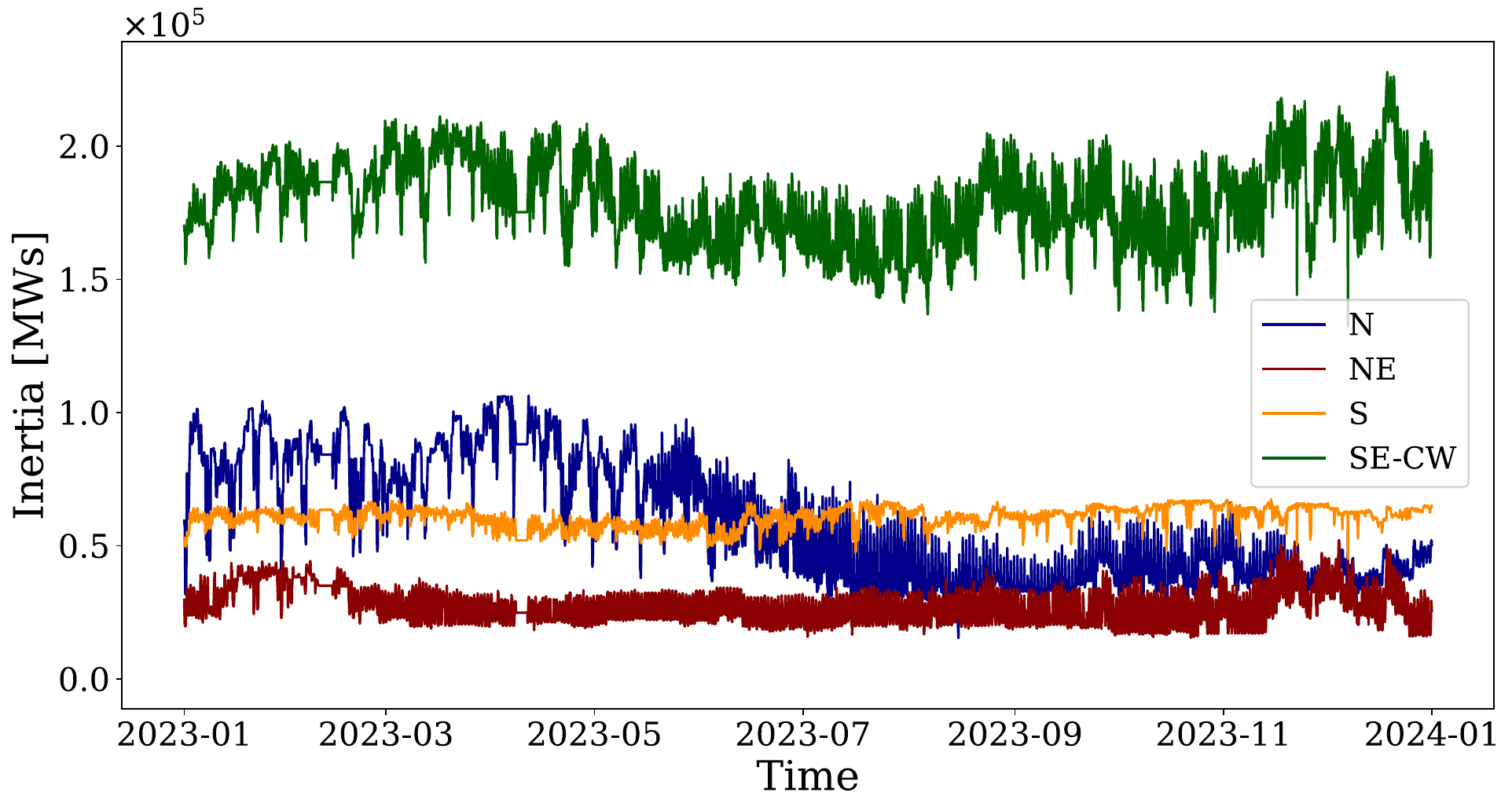}
\caption{Regional inertia estimation per Region in the BIPS.}
\label{fig:inertia}}
\end{figure}

Figure~\ref{fig:inertia} illustrates the estimated regional inertia for 2023. Throughout this year, the Northeast (NE) region exhibits the lowest regional inertia due to the high local penetration of IBR and lower dispatch of conventional generation. In contrast, the Southeast-Central West (SE-CW) region exhibits the highest regional inertia, which aligns with its status as the region with the largest synchronous generation capacity. The average regional inertia for each region is presented in Table~\ref{tab:inertia_horizontal}. In Section~\ref{sec:regional}, we explore the relationship between regional frequency fluctuations and their respective regional inertia. Given its distinctive characteristics of high IBR penetration and low inertia, the Northeast (NE) region is chosen as the primary focus for analyzing regional frequency dynamics.

\begin{table}[htb]
    \centering
    \caption{Average regional inertia values in the BIPS in 2023.}
    \label{tab:inertia_horizontal}
    \begin{tabular}{lcccc}
        \toprule
        \textbf{Region} & NE & N & S & SE-CW \\
        \midrule
        \textbf{Inertia} [MWs] & 27198 & 58318  & 60586 & 178207 \\
        \bottomrule
    \end{tabular}
\end{table}

\section{System-wide frequency evaluation}

This section evaluates the impact of IBR on the system-wide frequency dynamic of the BIPS. We consider the frequency measurements recorded from 2011 to 2023. Additionally, an evaluation of net load ramps and wind curtailment in the BIPS is presented. The wind curtailment data used in this study is publicly available in~\cite{dado_ons_open}.

\subsection{System-wide frequency evaluation}

In this analysis, the variation in the system-wide frequency is assessed using data collected from a PMU installed in Florianópolis, located in the southern region of the BIPS. The choice of a specific PMU is justified as we are analyzing the variation in system-wide frequency over an entire year. The impact of local variations is minimized when data from extended periods is considered, making the chosen PMU representative of overall system frequency trends. The PMU frequency measurements are downsampled with a sampling rate of 1 Hz to make it possible to process this large amount of data.

\begin{figure}[t]
{\centering
\includegraphics[width=1\linewidth]{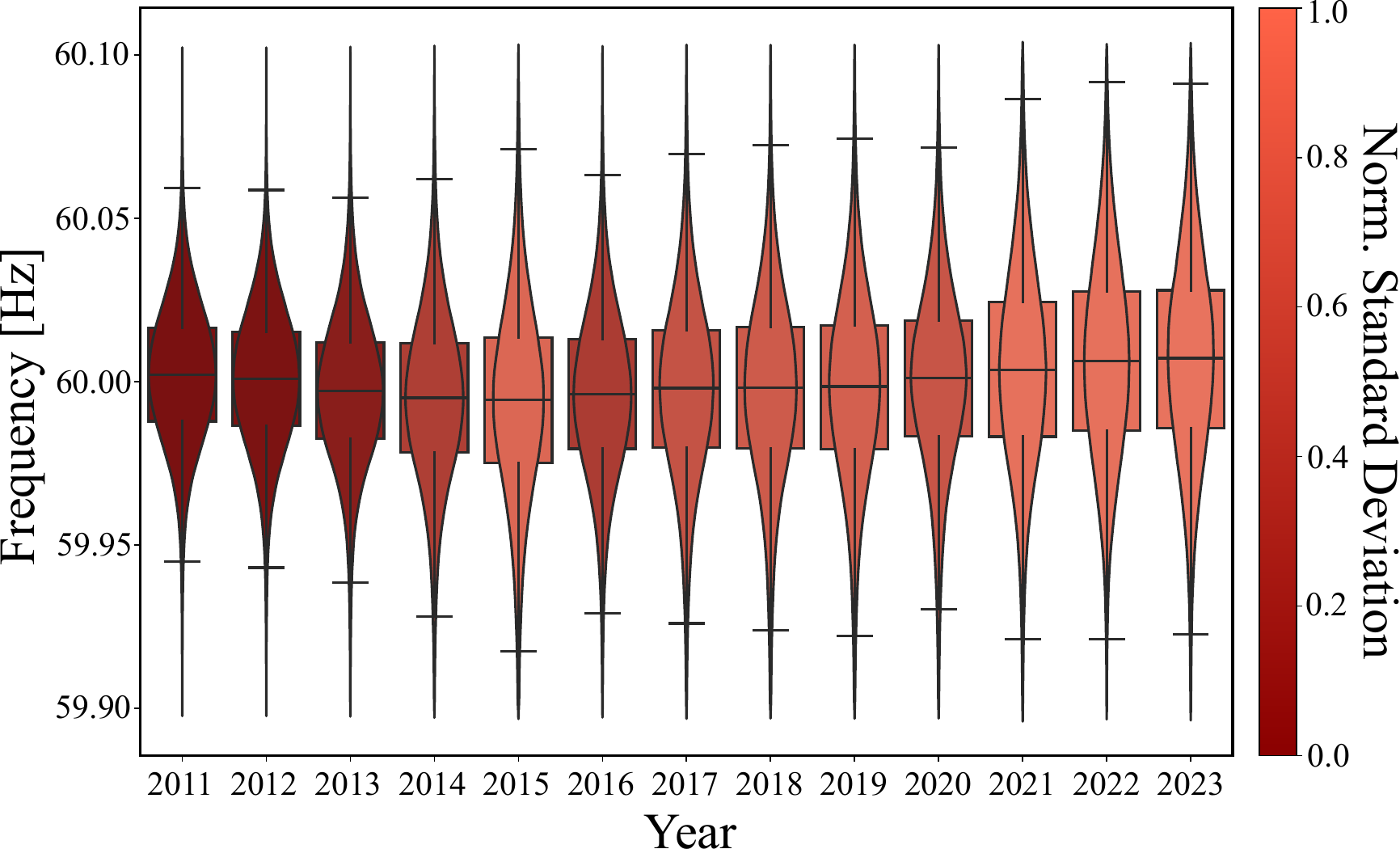}
\caption{Frequency variations and its normalized standard deviation in the BIPS from 2011 to 2023.}
\label{fig:freq_years}}
\end{figure}

The primary metric for this evaluation is the normalized standard deviation of the system-wide frequency, analyzed over multiple years. Figure~\ref{fig:freq_years} illustrates the frequency variation in the BIPS from 2011 to 2023 during the period between 09:00 AM and 11:00 AM, a timeframe characterized by low load demand and high IBR penetration. The normalized standard deviation of the frequency is depicted as a color scale. The results highlight two distinct phases in the evolution of frequency variations. The first phase, from 2011 to 2015, exhibits a decline in the average system-wide frequency, while the second phase, starting in 2016, reveals an increasing trend in average frequency.

\begin{figure}[tb]
{\centering
\includegraphics[width=0.95\linewidth]{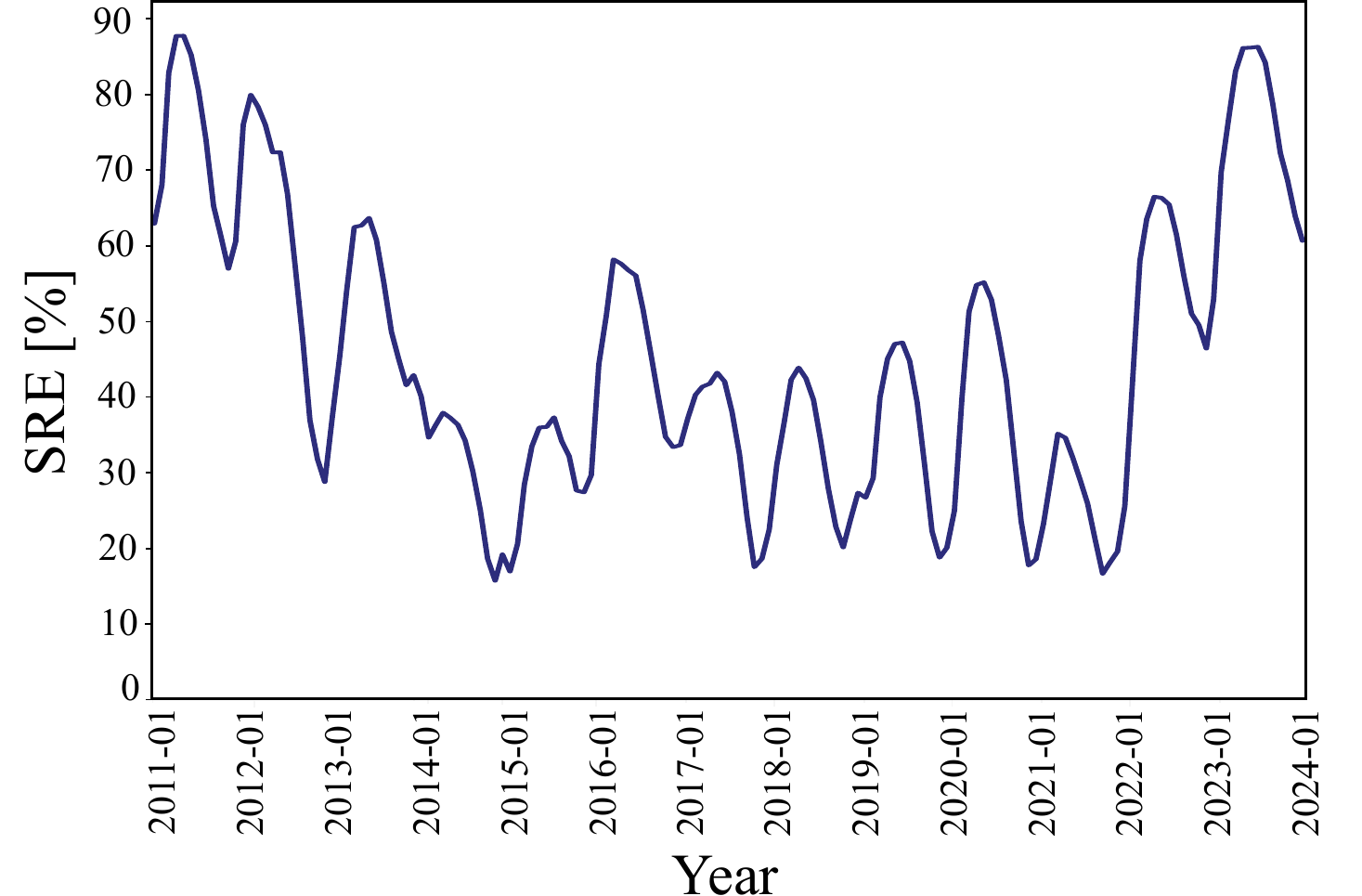}
\caption{Stored Reservoir Energy from 2011 to 2023 in the SE-CW region.}
\label{fig:era}}
\end{figure}

In the first phase, the system-wide frequency trend aligns with a period of significant water scarcity in the BIPS, as illustrated by the stored reservoir energy (SRE) in the SE-CW region shown in Figure~\ref{fig:era}. Hydroelectric generation, the main energy source for the Brazilian system, struggled to meet demand. For instance, in 2014, the hydroelectric generation accounted for 60\% of the generation matrix, while IBR contributed only 6\%. This shortfall led to challenges in allocating the spinning reserve and, as a consequence, more frequent underfrequency events.

\begin{figure}[tb]
{\centering
\includegraphics[width=1\linewidth]{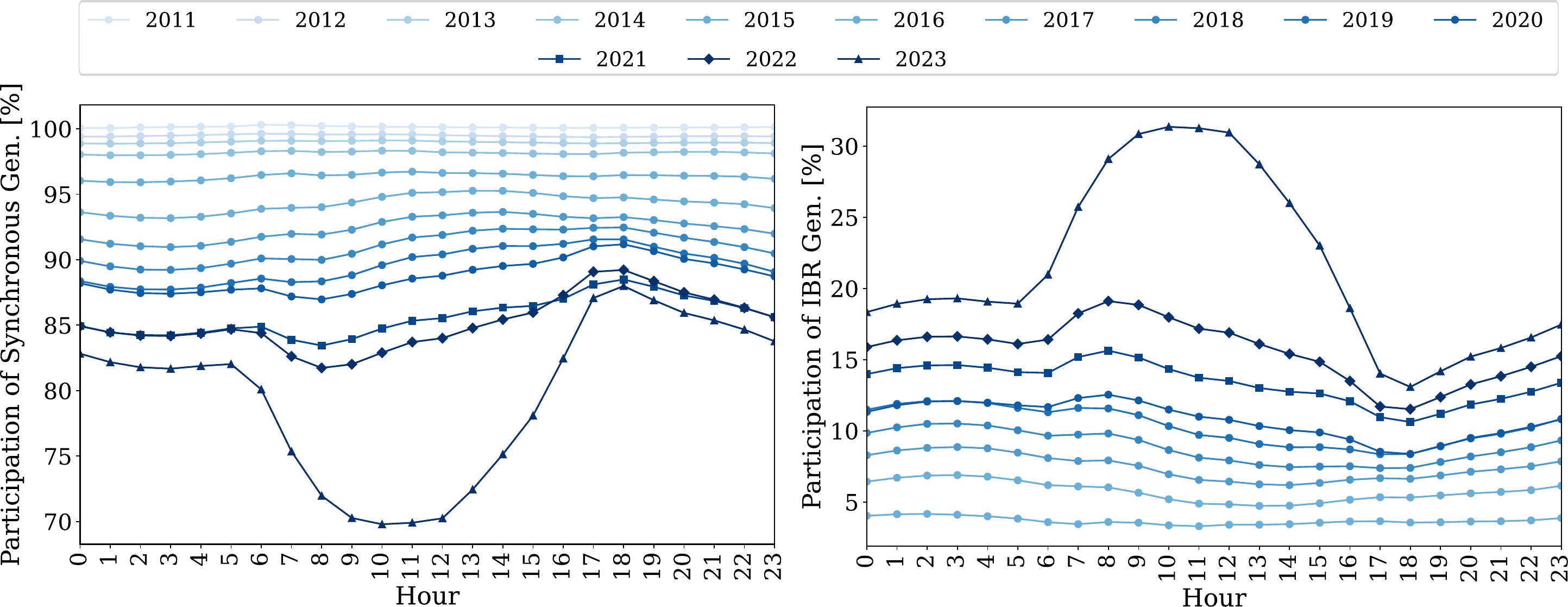}
\caption{Participation of synchronos and IBR generation to supply the demand.}
\label{fig:synpen}}
\end{figure}

In the second phase, from 2016 to 2023, the average frequency shows a steady increase over the years, but its variability also rises significantly. This period is marked by a declining contribution of synchronous generation (hydro and thermal) to meet demand, driven by the increasing penetration of IBR generation, as illustrated in Figure~\ref{fig:synpen}. 

Figure~\ref{fig:synpen} highlights a more pronounced difference in the contributions of IBR and synchronous generation to demand in 2023 compared to previous years. Two main factors explain this shift: (i) the increased penetration of IBR at both the transmission and distribution levels in 2023, resulting in a more substantial share of IBR in meeting demand; and (ii) the Brazilian system operator (ONS) improving its estimation of DER. Prior to 2023, DER contributions were not directly accounted for and were only reflected indirectly in the net demand at the transmission level. The enhanced DER accounting in 2023 provides a more accurate representation of IBR participation in the system, making direct comparisons with previous years more challenging.

As IBR generation continues to expand, the system's frequency response has been steadily deteriorating, particularly during periods of high IBR penetration and low demand, such as in the early morning hours. This reduction in synchronous generation has led to greater frequency volatility in the BIPS, as evidenced by the pronounced rise in the normalized standard deviation during the last three years of evaluation (2021-2023). Interestingly, while the SRE levels in 2011 and 2023 are comparable (see Figure~\ref{fig:era}), the frequency dynamics in 2023 exhibit greater variability, underscoring the impact of high IBR penetration. Moreover, the limited frequency regulation capabilities of wind and solar generation—offering minimal to no support for frequency control— may have further amplified these variations.

The observed increase in the average frequency can be attributed to surplus generation during periods of low load demand, driven by the diversification of the Brazilian energy matrix during the second phase, particularly through the expansion of concentrated wind and distributed solar generation. This diversification has introduced new challenges, as evidenced by the growing net load ramps and the curtailment of wind generation. These aspects are explored in detail in the following subsection.

\subsection{Net load ramps and Wind Curtailment}

The net load is the key metric used to quantify the amount of load that conventional generation must supply to maintain the balance between generation and demand, ensuring frequency stability. The rise in net load ramps over recent years has likely contributed to system-wide frequency variations The net load is calculated using~\eqref{eq:netload}, while net load ramps are defined as changes in net load over a specified time window, $h$, calculated using~\eqref{eq:netloadramp}. In this study, net load ramps are evaluated for $h = \{1, 3, 5, 7, 12\}$ hours. 
\begin{equation}
\label{eq:netload}
    NL = P_{\text{L}} - P_{\text{wind}} - P_{\text{solar}} - P_{\text{DER}}
\end{equation}
\begin{equation}
\label{eq:netloadramp}
    R = NL_{i+h} - NL_i
\end{equation}

\begin{figure}[!htb]
{\centering
\includegraphics[width=0.85\linewidth]{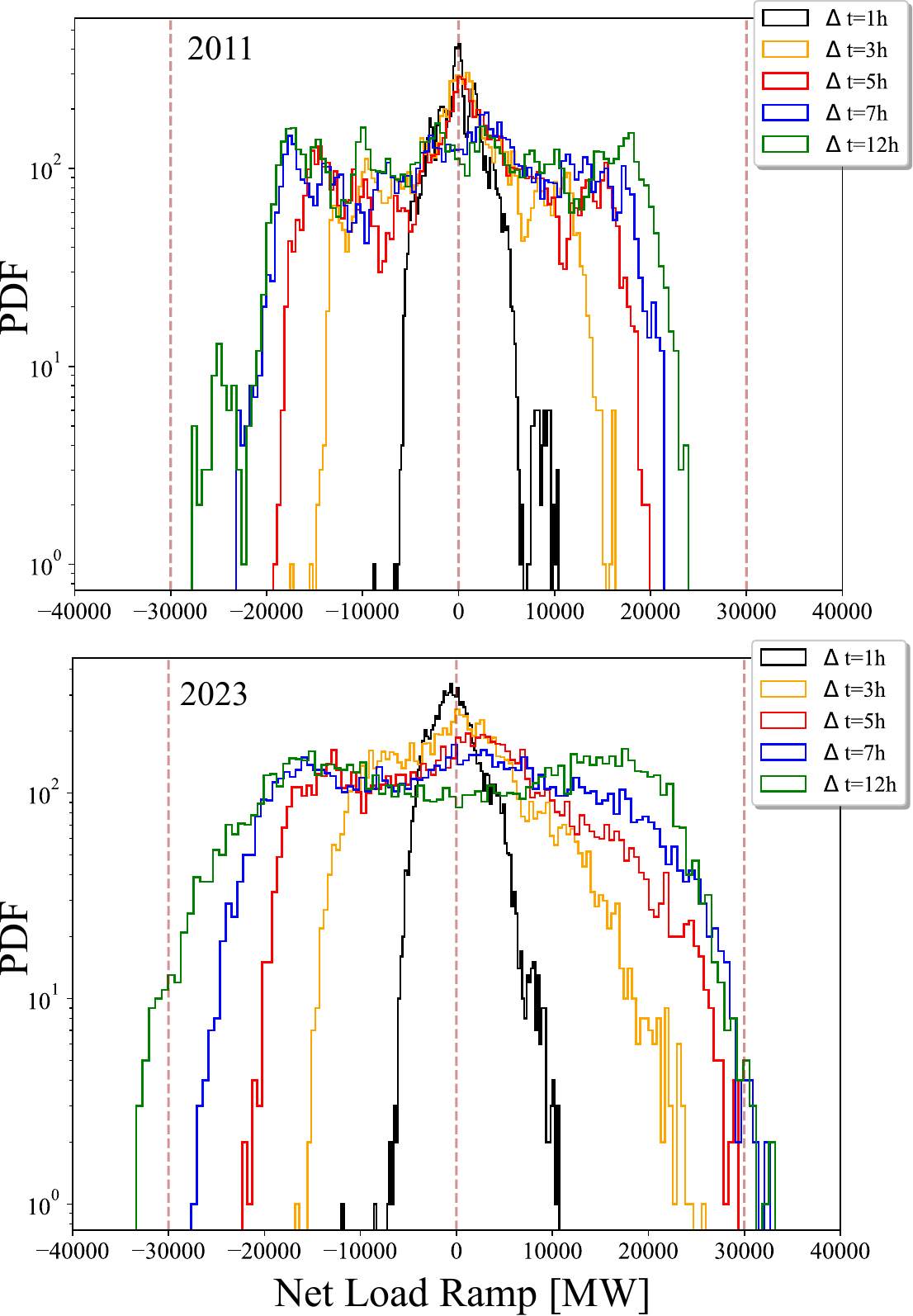}
\caption{Net load ramps in the BIPS for 2011 and 2023 for different time intervals.}
\label{fig:netloadramps2023}}
\end{figure}

Figure~\ref{fig:netloadramps2023} compares the net load ramps for 2011 (top) and 2023 (bottom). A noticeable increase in both the magnitude and frequency of these ramps is evident, particularly for positive net load ramps. In 2023, there is a marked rise in the magnitude of positive net load ramps over time windows of 3, 5, 7, and 12 hours. For 1-hour ramps, an increased frequency of positive ramps around 10 GW is observed. Additionally, for 3, 5, 7, and 12-hour ramps, a significant increase in values around 30 GW is observed in 2023. Regarding negative net load ramps, the most substantial increases occur over 7- and 12-hour time windows, with values reaching approximately -30 GW.

To better understand how the rise in wind generation has influenced positive net load ramps and contributed to surplus generation in 2023, wind curtailment data is analyzed. Curtailment refers to the reduction in generation output imposed by the system operator, despite the availability of potential energy. Curtailment occurs due to various reasons, including transmission limitations, reliability concerns, or system balancing challenges~\cite{curtailment_rev}. In the context of the BIPS, it is particularly important to note that the system operator exercises greater control over wind generation plants compared to solar generation, as most solar generation is distributed rather than centrally managed.

\begin{figure}[tb]
{\centering
\includegraphics[width=1\linewidth]{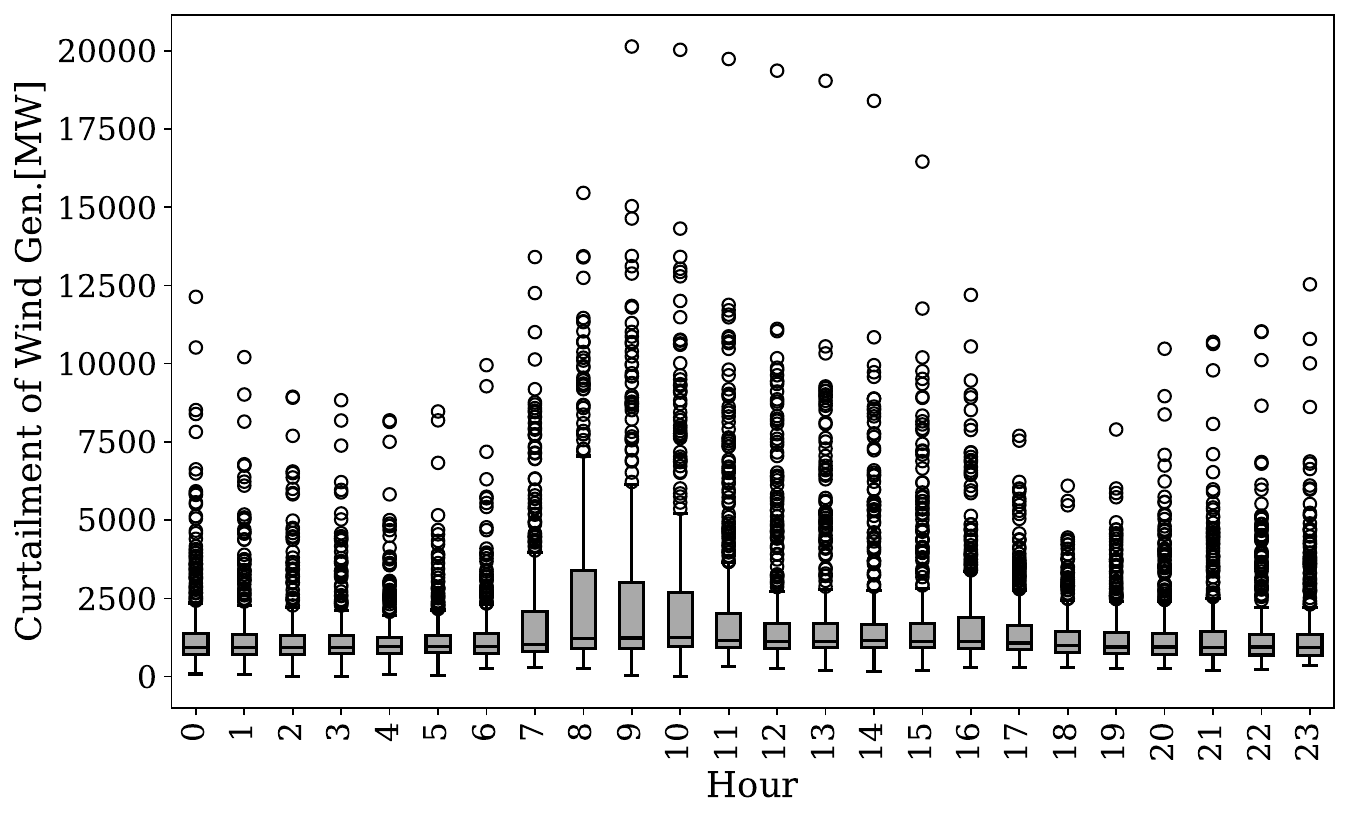}
\caption{Total curtailment of wind generation in 2023 per hour.}
\label{fig:co_wind}}
\end{figure}

Figure~\ref{fig:co_wind} presents the hourly distribution of wind generation curtailment in 2023. The highest levels of curtailment are observed between 8:00 AM and 11:00 AM, a period characterized by reduced demand and high wind generation capacity. The reasons for wind curtailment during this period are summarized in Table~\ref{tab:reasons_stats}. Notably, energy balance-related constraints, which reflect the need to balance generation and demand, account for over 50\% of curtailment during these period of the day. Reliability requirements and external unavailability (electrical) contribute to the remaining share, highlighting the multifaceted challenges of integrating high levels of wind generation.

\begin{table}[htb]
    \centering
    \caption{Reasons for curtailment during the period from 8 AM to 11 AM.}
    \begin{tabular}{lcc}
        \toprule
        \textbf{Reason} & \textbf{Count} & \textbf{Percentage [\%]} \\
        \midrule
        Energy balance & 39,270 & 51.85 \\
        Reliability Requirements & 19,650 & 25.94 \\
        External Unavailability (Electrical) & 16,818 & 22.21 \\
        \bottomrule
    \end{tabular}
    \label{tab:reasons_stats}
\end{table}

Therefore, the observed trend of increasing positive net load ramps, coupled with the high levels of wind generation during the 08:00 AM to 11:00 AM period, provides a clear explanation for the consistent increase in the average system-wide frequency, as illustrated in Figure~\ref{fig:freq_years}. This relationship highlights the growing influence of IBR on system-wide frequency variations in the BIPS.

\section{Local frequency fluctuation evaluation}
\label{sec:regional}

\begin{figure}[htb]
{\centering
\includegraphics[width=1\linewidth]{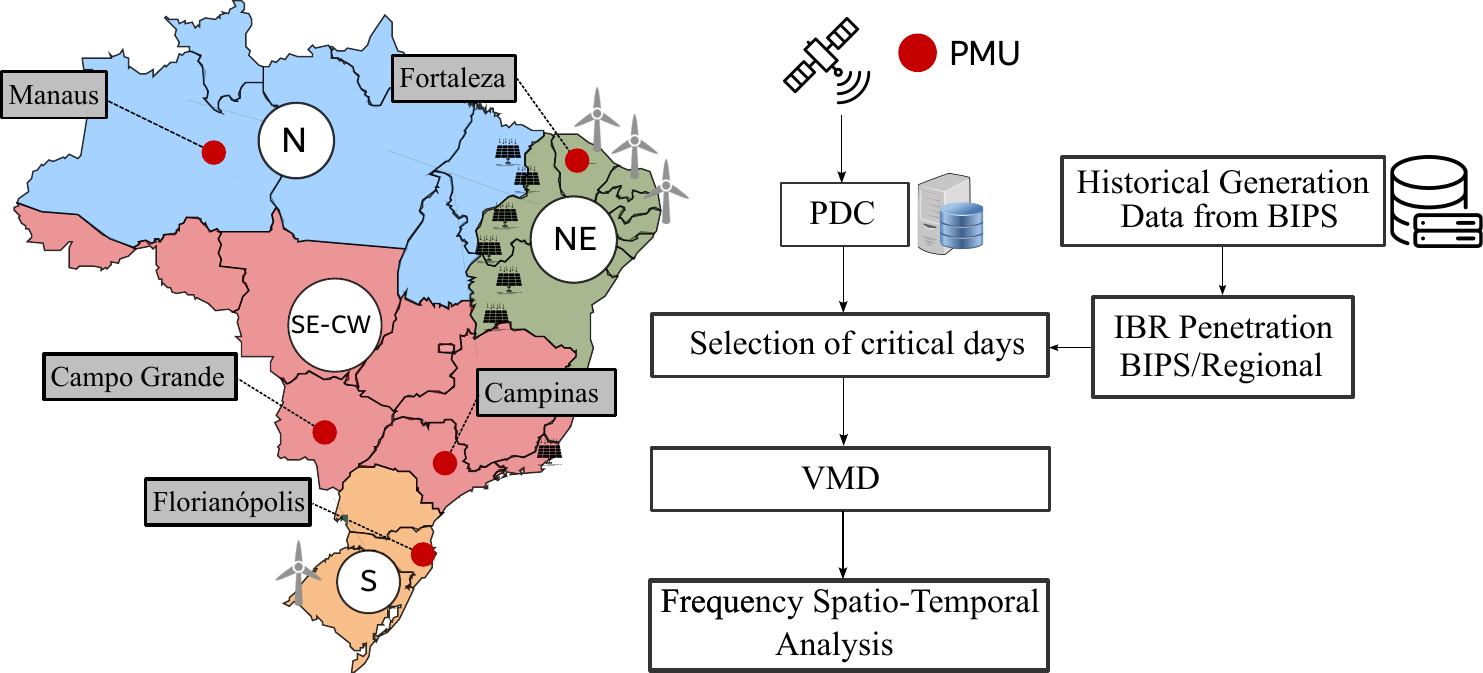}
\caption{Framework for regional frequency response evaluation}
\label{fig:PMU_Framework}}
\end{figure}

The proposed methodology for evaluating local frequency fluctuations is outlined in Figure~\ref{fig:PMU_Framework}. The methodology comprises four main steps: (i) calculation of local IBR penetration, (ii) selection of critical week, (iii) extraction of local frequency fluctuations, and (iv) dynamic frequency evaluation.

Five PMUs are selected to evaluate the regional frequency dynamic during 2023: Fortaleza (NE), Manaus (N), Campo Grande (CW), Campinas (SE), and Florianópolis (S). In order to properly track the local dynamic behavior of the regions, the frequency data are with a sampling rate of 30Hz.

\subsection{Week with highest IBR penetration}

\begin{figure}[htb]
{\centering
\includegraphics[width=1\linewidth]{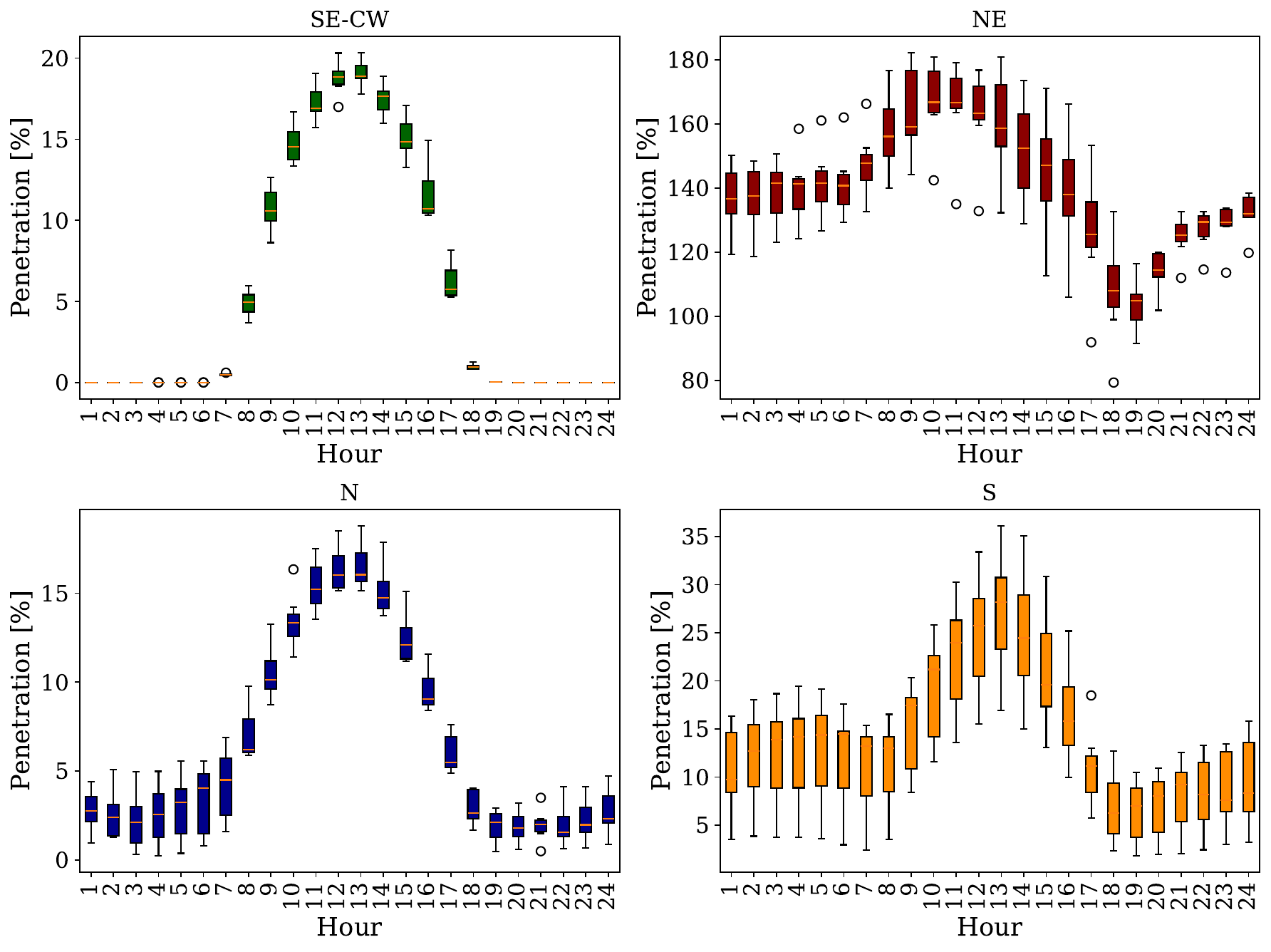}
\caption{IBR penetration by hour for each geo-electrical region for the critical week.}
\label{fig:regional_NE}}
\end{figure}

The critical week for frequency evaluation is selected based on the highest IBR penetration in the Northeast region. The week with the highest average IBR penetration in relation to regional demand is from 17/07/2023 to 23/07/2023. The average IBR penetration in the Northeast region during this week is about 140\%. In addition, to determine the hours for frequency evaluation and regional IBR impact assessment, the penetration of IBR for each hour in this period is presented in Figure~\ref{fig:regional_NE}. Based on this, two datasets are formed: 

\begin{itemize}
    \item Group I: The period with the highest regional IBR penetration in all geo-electric regions, which comprises frequency measurements between 09:00 and 12:00.
    \item Group II: The period with the lowest IBR penetration among all geo-electric regions, which covers the period from 21:00 to 00:00. We did not select times with lowest average penetration (17:00-19:00) to avoid potential variations due to solar and wind generation ramps. 
\end{itemize}

\subsection{Local frequency fluctuation evaluation}

The PMU frequency measurements of each region for the critical week were used in the VMD approach to extract the dynamic component of the frequency. 

\begin{figure}[!h]
{\centering
\includegraphics[width=0.85\linewidth]{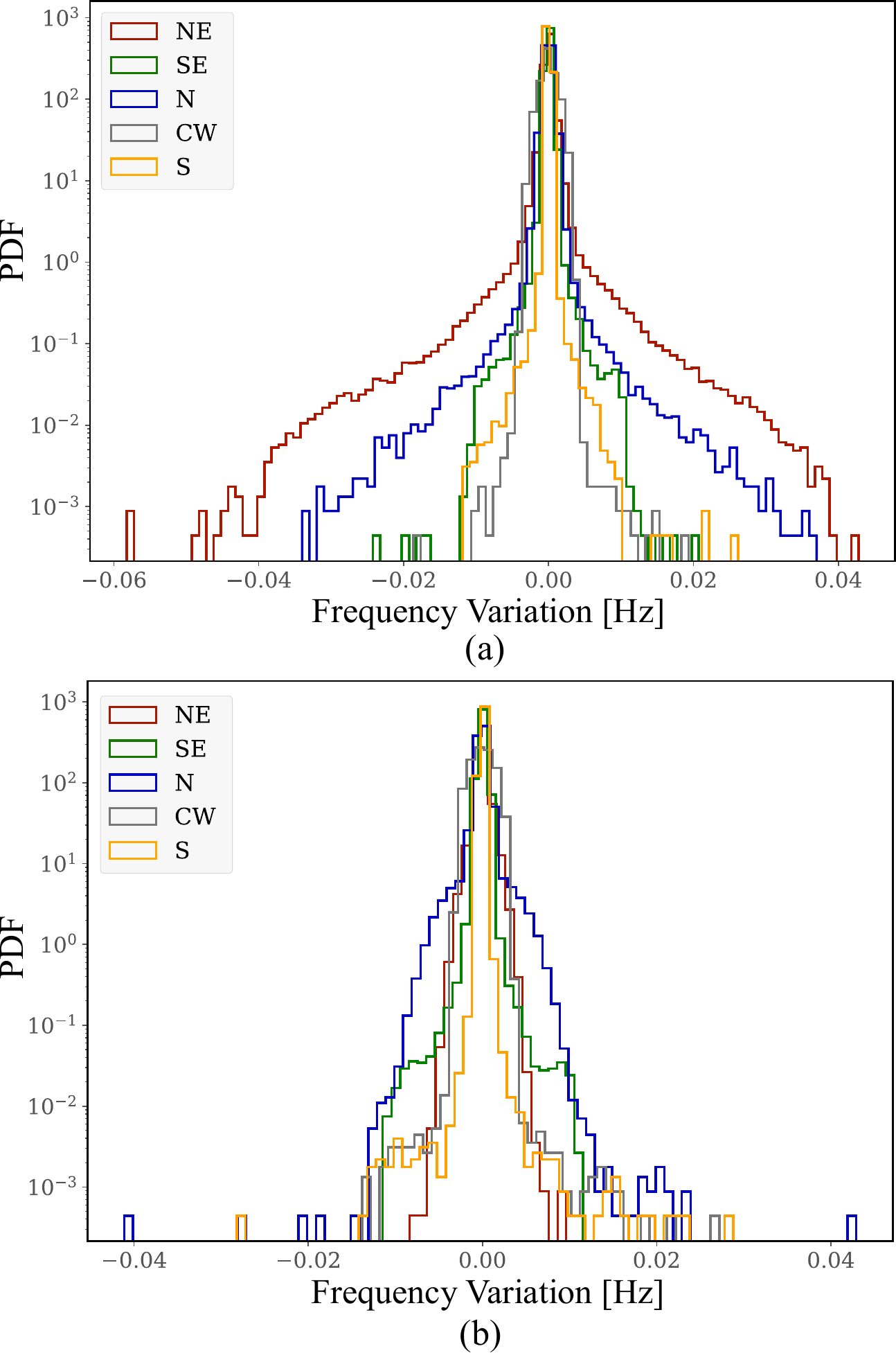}
\caption{Histogram of the local dynamic frequency behavior for (a) the period of highest IBR penetration in NE region (09:00 to 12:00) and (b) lowest IBR penetration in NE region (21:00 to 00:00).}
\label{fig:histogram_rocof_ab}}
\end{figure}

The histogram of the dynamic component of the regional frequency is evaluated considering the periods of highest and lowest IBR penetration. Figure~\ref{fig:histogram_rocof_ab}a shows the histogram of the dynamic component for the period with the highest penetration of IBR (Group I). The frequency variation in the Northeast (NE) region is the most pronounced, followed by the North (N) region. The remaining regions exhibit similar behavior with respect to the dispersion of the local frequency dynamics. Additionally, in Figure~\ref{fig:histogram_rocof_ab}a, we can notice an asymmetry in the frequency dispersion of the Northeast region. This asymmetry in frequency variation can be associated with the  non-linearity of the wind turbines, as discussed in~\cite{kerci2024asymmetry}.

Figure~\ref{fig:histogram_rocof_ab}b depicts the histogram for Group II, representing the period with the lowest penetration of IBR during the week of analysis. Notably, all regions present similar dispersion in their regional dynamic frequency behavior. This result highlights the impact of regional IBR penetration in the local frequency dispersion compared to the COI frequency.

\subsection{Correlation Analysis}

To evaluate the correlation between the increasing penetration of IBR, the regional inertia, and the frequency fluctuations in each region, we use Pearson's correlation coefficient. The Pearson's correlation coefficient provides a quantitative measure of the linear correlation between two variables and is calculated as follows:
\begin{equation}
    \label{eq:pearson}
    r = \frac{\sum_{i}^{N}  \left( x_i -\Bar{x} \right)\left( y_i -\Bar{y} \right)  }{\sqrt{ \sum_{i}^{N}  \left( x_i -\Bar{x} \right)^2 \sum_{i}^{N}  \left( y_i -\Bar{y} \right)^2  }},
\end{equation}
\noindent where $x_i$ and $y_i$ are the values of the two variables at time stamp $i$, $N$ is the total number of observations, and $\Bar{x}$ and $\Bar{y}$ are the mean values of the time series $x$ and $y$, respectively. A positive value of $r$ indicates a positive linear correlation, meaning that an increase in IBR penetration in a specific region results in an increase in the standard deviation of local frequency fluctuations. 

This analysis considers the complete frequency measurements for all days within the critical week. Since the inertia and IBR penetration data are available with one-hour resolution, we calculate the standard deviation ($\sigma_f$) of the local frequency variation over each one-hour period. Table~\ref{tab:combined_correlation} presents the resulting correlations between the standard deviation of local frequency in each region and the regional inertia and IBR penetration. The Northeast region shows a positive linear correlation between regional IBR penetration and the standard deviation of local frequency. Similarly, the Southeast exhibits a linear correlation between IBR penetration in the SE-CW area and its local frequency variation. For the remaining regions, the correlation is negligible. Furthermore, the Northeast displays the strongest negative correlation with its regional inertia, underscoring the significant impact of inertia on local frequency fluctuations.

\begin{table}[htb]
    \centering
    \caption{Correlation coefficients for inertia and IBR penetration with $\sigma_f$ for each region.}
    \label{tab:combined_correlation}
    \begin{tabular}{lcc}
        \toprule
        \textbf{Region} & $r_{\text{Inertia}}$ & $r_{\text{IBR}}$ \\
        \midrule
        NE & -0.162 & 0.258 \\
        S & -0.075 & 0.023 \\
        N & -0.115 & 0.004 \\
        CW & -0.035 & -0.035 \\
        SE & 0.156 & 0.105 \\
        \bottomrule
    \end{tabular}
\end{table}

It is important to note that correlations close to one were not observed. Two factors can be highlighted: (i) the use of regional inertia values measured at hourly intervals may not accurately capture the average inertia over the entire one-hour window, and (ii) the inertia values for each geo-electric region may not fully represent local frequency behavior, potentially requiring nodal inertia measurements~\cite{Bruno2023,Fanhong2020} or regional inertia data from more localized areas near the measurement points~\cite{Misyris2023}.

\subsection{Autocorrelation Analysis}

Similarly, the autocorrelation of the frequency fluctuation can be evaluated. Autocorrelation function (ACF) measures the similarity between a signal and a time-shifted version of itself by a lag $\tau$. For a given lag 
$\tau$, the autocorrelation can be computed using the Pearson correlation coefficient~\eqref{eq:pearson}, with 
$y$ representing the shifted signal. 

\begin{figure}[!htb]
{\centering
\includegraphics[width=1\linewidth]{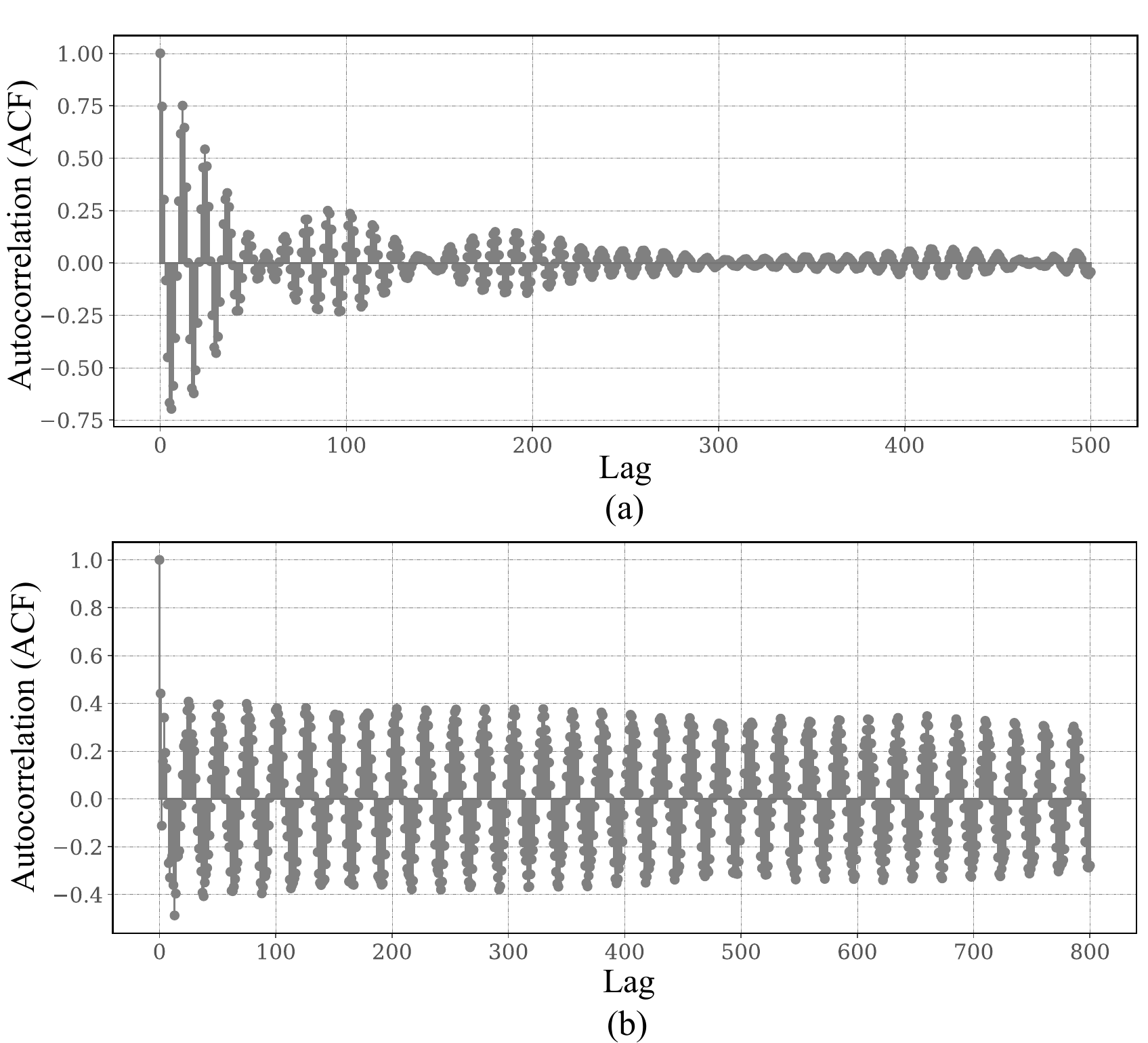}
\caption{Autocorrelation of the dynamic component for (a) Group (I) with higher local IBR penetration and (b) for Group (II) with lower IBR penetration.}
\label{fig:ACF}}
\end{figure}

The ACF of the frequency dynamic component is presented in Figure~\ref{fig:ACF}. Figure~\ref{fig:ACF}a shows the ACF obtained for Group I, where Figure~\ref{fig:ACF}b presents the ACF for Group II. It is note that the dynamic frequency component for the case with the highest IBR penetration present a more stochastic behavior, since the ACF tends to zero for lags greater than 1. On the other hand, the case with the lowest IBR penetration is with a periodic ACF, which can indicate that the dynamic frequency component is more related to the system oscillations. Therefore, the increase penetration of IBR introduce stochastic variations in the local frequency dynamic, directly impacting the local frequency variations.

\subsection{Converter-driven oscillation event}

\begin{figure}[!htb]
{\centering
\includegraphics[width=1\linewidth]{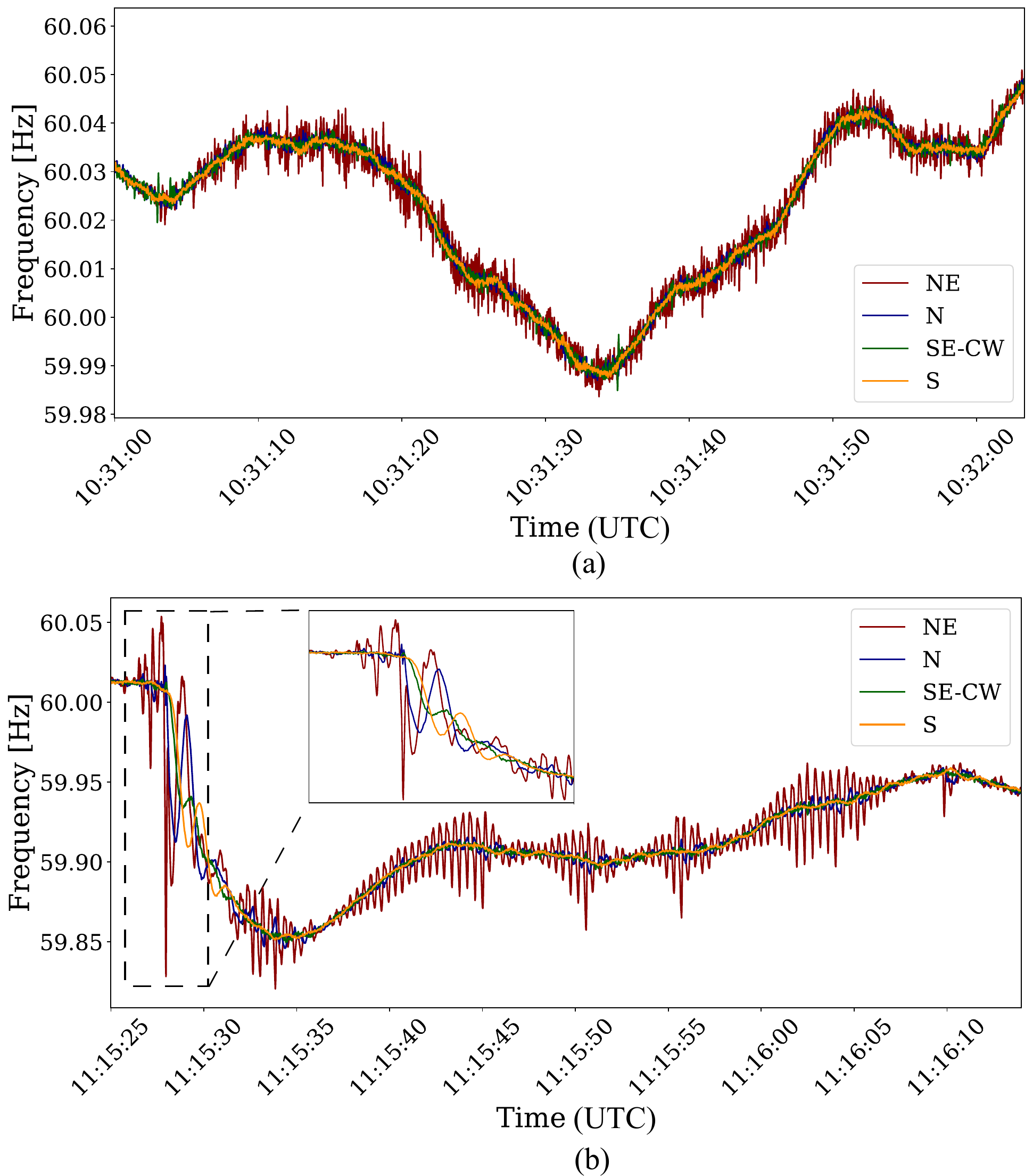}
\caption{Frequency variation in selected PMUs of different geo-electrical region after an event in 20 of July of 2023.}
\label{fig:disturbio}}
\end{figure}

This section highlights the impact of converter-driven oscillation observed in the BIPS in 2023. On July 20th, during the critical week, 2.5 Hz oscillations were observed in the Northeast region under normal operating conditions and following a disturbance. Figure~\ref{fig:disturbio}a depicts the frequency in each region prior to the event, with the Northeast showing greater variability during normal operations. Using the Fast Fourier Transform (FFT) analysis during ambient conditions, we can identify the presence of 2.5 Hz oscillations. These oscillations were particularly evident in the voltage magnitude and frequency.

Following the automatic shutdown of a transformer, over 500 MW of wind generation was curtailed, resulting in the frequency drop depicted in Figure~\ref{fig:disturbio}b. The 2.5 Hz oscillations intensified after the event, with the Northeast region experiencing the highest local RoCoF. 

During this event, the wind generation reached 18 GW in the NE region and the synchronous generation 2 GW, highlighting the potential influence of IBR on the observed oscillations. PMU data from the Northeast transmission system revealed that the voltage oscillations were more pronounced near certain solar power plants. Attempts to mitigate the oscillations showed that they diminished after the disconnection of solar generation from that region.

These observations suggest that the oscillations were likely driven by interactions between IBR and the system, possibly under conditions of system weakness. The high IBR penetration in the region, coupled with the localized impact of solar power plants and high frequency oscillation, supports the hypothesis of converter-driven oscillations. Further analysis of system conditions and IBR behavior is needed to confirm the specific mechanisms driving this phenomenon.

\section{Discussion}

The evaluation of the systemic frequency behavior in the BIPS revealed a clear trend toward overfrequency during periods of high IBR penetration over recent years (2015-2023). The consistent growth of IBR has significantly contributed to generation oversupply. Alongside this increase in overfrequency, a rise in frequency volatility has also been observed in both local and systemic frequency. Based on these findings, the following key points merit discussion:

\begin{itemize}
    
    \item \textbf{Addressing Frequency Variability}: The increasing variability in both systemic and local frequencies calls for a reassessment of acceptable limits for frequency fluctuations to maintain the dynamic stability of the system. Studies in the literature recommend adopting or enhancing the Automatic Generation Control (AGC) to reduce frequency variability~\cite{Australia_TPWRS,Kerci2023}. Currently, the BIPS employs AGC within each geo-electric area. In this context, further advancements may be required to optimize the placement of reserves of generators that participate in the AGC. 
    
    \item \textbf{Improving Regional Inertia Assessment}: With the increasing local frequency variability, there is a critical need to enhance the methodology for assessing regional inertia within the system. The current approach estimates inertia by region based on dispatch information from synchronous machines. A more effective method would involve using PMU-based measurement data to estimate effective inertia—encompassing contributions from synchronous generators, loads, and synthetic inertia. This approach is better suited to the evolving characteristics of the BIPS.
\end{itemize}

\section{Conclusion}

This paper utilizes real-world PMU data from BIPS, a large-scale, meshed power system, to assess the impact of increasing IBR penetration on dynamic frequency response. The system-wide frequency has shown a gradual degradation, coinciding with an increase in the net load ramps. Additionally, we analyzed the effect of regional IBR penetration on regional frequency dynamics. A methodology based on VMD is proposed to assess local frequency fluctuations at different regions within BIPS. A critical week in 2023, marked by the highest IBR penetration, was evaluated, revealing that the frequency in the Northeast region is particularly sensitive to IBR penetration. These findings highlight the need for more regional and localized evaluations of frequency stability metrics, such as regional inertia estimation. Future work will focus on determining critical regional inertia values in the BIPS and evaluating new metrics for regional frequency stability assessment. Additionally, further investigation will be conducted on optimal parameter selection for VMD and the decomposition of COI and local frequency variations.

\section*{Acknowledgement}

We would like to thank Rafael Zymler from ONS (the Brazilian SO) for providing the historical inertia data. This work was supported in part by Brazilian Agency Coordination for the Improvement of Higher Education Personnel (CAPES) (Finance Code 001) and by FAPESP (grant 2021/11380-5). This work is also supported by Engie under the Research and Development Program regulated by the Brazilian Electricity Regulatory Agency (ANEEL) (PD-00403-0053/202).

\bibliographystyle{bibliography/IEEEtran}
\bibliography{bibliography/IEEEabrv,./bibliography/refs}

\begin{thebibliography}{10}
\providecommand{\url}[1]{#1}
\csname url@samestyle\endcsname
\providecommand{\newblock}{\relax}
\providecommand{\bibinfo}[2]{#2}
\providecommand{\BIBentrySTDinterwordspacing}{\spaceskip=0pt\relax}
\providecommand{\BIBentryALTinterwordstretchfactor}{4}
\providecommand{\BIBentryALTinterwordspacing}{\spaceskip=\fontdimen2\font plus
\BIBentryALTinterwordstretchfactor\fontdimen3\font minus \fontdimen4\font\relax}
\providecommand{\BIBforeignlanguage}[2]{{%
\expandafter\ifx\csname l@#1\endcsname\relax
\typeout{** WARNING: IEEEtran.bst: No hyphenation pattern has been}%
\typeout{** loaded for the language `#1'. Using the pattern for}%
\typeout{** the default language instead.}%
\else
\language=\csname l@#1\endcsname
\fi
#2}}
\providecommand{\BIBdecl}{\relax}
\BIBdecl

\bibitem{RAP}
\BIBentryALTinterwordspacing
``Relatório de análise de pertubação - rap: Análise da perturbação do dia 15/008/2023 às 08h30min,'' accessed on June, 2024. [Online]. Available: \url{https://bit.ly/4eaW2if}
\BIBentrySTDinterwordspacing

\bibitem{AEMC2020}
AEMC, ``Mandatory primary frequency response, rule determination,'' 26 March 2020.

\bibitem{Australia_TPWRS}
H.~H. Alhelou, B.~Bahrani, J.~Ma, and D.~J. Hill, ``Australia's power system frequency: Current situation, industrial challenges, efforts, and future research directions,'' \emph{IEEE Transactions on Power Systems}, pp. 1--13, 2023.

\bibitem{Cai2024}
X.~Cai, N.~Zhang, E.~Du, Z.~An, N.~Wei, and C.~Kang, ``Low inertia power system planning considering frequency quality under high penetration of renewable energy,'' \emph{IEEE Transactions on Power Systems}, vol.~39, no.~2, pp. 4537--4548, 2024.

\bibitem{Kerci2023}
T.~Kerçi, M.~Hurtado, M.~Gjergji, S.~Tweed, E.~Kennedy, and F.~Milano, ``Frequency quality in low-inertia power systems,'' in \emph{2023 IEEE Power \& Energy Society General Meeting (PESGM)}, 2023, pp. 1--5.

\bibitem{Xypolytou2018}
\BIBentryALTinterwordspacing
E.~Xypolytou, W.~Gawlik, T.~Zseby, and J.~Fabini, ``Impact of asynchronous renewable generation infeed on grid frequency: Analysis based on synchrophasor measurements,'' \emph{Sustainability}, vol.~10, no.~5, 2018. [Online]. Available: \url{https://www.mdpi.com/2071-1050/10/5/1605}
\BIBentrySTDinterwordspacing

\bibitem{Adeen2019}
M.~Adeen, G.~M. Jónsdóttir, and F.~Milano, ``Statistical correlation between wind penetration and grid frequency variations in the irish network,'' in \emph{2019 IEEE International Conference on Environment and Electrical Engineering and 2019 IEEE Industrial and Commercial Power Systems Europe (EEEIC / I\&CPS Europe)}, 2019, pp. 1--6.

\bibitem{Milano_FDF}
F.~Milano and A.~Ortega, ``Frequency divider,'' \emph{IEEE Transactions on Power Systems}, vol.~32, no.~2, pp. 1493--1501, 2017.

\bibitem{Giudice2021}
D.~{del Giudice}, A.~Brambilla, S.~Grillo, and F.~Bizzarri, ``Effects of inertia, load damping and dead-bands on frequency histograms and frequency control of power systems,'' \emph{International Journal of Electrical Power \& Energy Systems}, vol. 129, p. 106842, 2021.

\bibitem{Francesca2016}
F.~M. Mele, A.~Ortega, R.~Zarate-Minano, and F.~Milano, ``Impact of variability, uncertainty and frequency regulation on power system frequency distribution,'' in \emph{2016 Power Systems Computation Conference (PSCC)}, 2016, pp. 1--8.

\bibitem{Zuo2021}
Y.~Zuo, Z.~Yuan, F.~Sossan, A.~Zecchino, R.~Cherkaoui, and M.~Paolone, ``Performance assessment of grid-forming and grid-following converter-interfaced battery energy storage systems on frequency regulation in low-inertia power grids,'' \emph{Sustainable Energy, Grids and Networks}, vol.~27, p. 100496, 2021.

\bibitem{Vorobev2019}
P.~Vorobev, D.~M. Greenwood, J.~H. Bell, J.~W. Bialek, P.~C. Taylor, and K.~Turitsyn, ``Deadbands, droop, and inertia impact on power system frequency distribution,'' \emph{IEEE Transactions on Power Systems}, vol.~34, no.~4, pp. 3098--3108, 2019.

\bibitem{Bruno2023}
B.~Pinheiro, L.~Viola, J.~H. Chow, and D.~Dotta, ``An analytical formulation for mapping the spatial distribution of nodal inertia in power systems,'' \emph{IEEE Access}, vol.~11, pp. 45\,364--45\,376, 2023.

\bibitem{Jiaxin2024}
J.~Wang, J.~Zhang, Q.~Hou, H.~Jiang, G.~Strbac, F.~Teng, and N.~Zhang, ``Analytical decomposition of nodal frequency responses,'' \emph{IEEE Transactions on Power Systems}, pp. 1--4, 2024.

\bibitem{Licheng2024}
L.~Wang, J.~Ren, G.~Huang, L.~Xie, C.~Feng, and Y.~Zhang, ``Identifying the largest rocof and its implications,'' \emph{IEEE Transactions on Power Systems}, pp. 1--4, 2024.

\bibitem{Guilherme2024}
G.~Albino, B.~Pinheiro, D.~Sanchez, and D.~Dotta, ``A framework for multiple-scenarios stability evaluation for the brazilian interconnected power system,'' in \emph{2024 4th International Conference on Smart Grid and Renewable Energy (SGRE)}, 2024, pp. 1--6.

\bibitem{David2024}
D.~Sanchez, B.~Pinheiro, L.~Lugnani, A.~F.~C. Aquino, and D.~Dotta, ``Multiple-case voltage profile evaluation of brazilian interconnected power system,'' in \emph{2024 IEEE Power \& Energy Society General Meeting (PESGM)}, 2024, pp. 1--5.

\bibitem{Guido2022}
G.~R. Moraes, B.~A. Ambrósio, J.~L. Pereira, D.~Issicaba, A.~F. Aquino, and I.~C. Decker, ``Impact analysis of covid-19 pandemic on the electricity demand, frequency control and electromechanical oscillation modes of the brazilian interconnected power system using low voltage wams data,'' \emph{International Journal of Electrical Power \& Energy Systems}, vol. 142, p. 108266, 2022.

\bibitem{lingling2023}
L.~Fan, Z.~Miao, and Z.~Wang, ``A new type of weak grid ibr oscillations,'' \emph{IEEE Transactions on Power Systems}, vol.~38, no.~1, pp. 988--991, 2023.

\bibitem{Osipov2023}
D.~Osipov, S.~Konstantinopoulos, and J.~H. Chow, ``A cross-power spectral density method for locating oscillation sources using synchrophasor measurements,'' \emph{IEEE Transactions on Power Systems}, vol.~38, no.~6, pp. 5526--5534, 2023.

\bibitem{Mario2019}
M.~R. {Arrieta Paternina}, R.~K. Tripathy, A.~Zamora-Mendez, and D.~Dotta, ``Identification of electromechanical oscillatory modes based on variational mode decomposition,'' \emph{Electric Power Systems Research}, vol. 167, pp. 71--85, 2019.

\bibitem{Zamora2019}
A.~Zamora, D.~Dotta, J.~H. Chow, R.~Tripathy, and M.~Paternina, ``Data-driven modal features extraction through the variational mode decomposition method,'' in \emph{2019 IEEE PES Innovative Smart Grid Technologies Conference - Latin America (ISGT Latin America)}, 2019, pp. 1--5.

\bibitem{Dragomiretskiy2014}
K.~Dragomiretskiy and D.~Zosso, ``Variational mode decomposition,'' \emph{IEEE Transactions on Signal Processing}, vol.~62, no.~3, pp. 531--544, 2014.

\bibitem{Carvalho2020}
V.~R. Carvalho, M.~F. Moraes, A.~P. Braga, and E.~M. Mendes, ``Evaluating five different adaptive decomposition methods for eeg signal seizure detection and classification,'' \emph{Biomedical Signal Processing and Control}, vol.~62, p. 102073, 2020.

\bibitem{ONS_CAPACITY}
\BIBentryALTinterwordspacing
\emph{Brazilian System Operator. (2023). The system in numbers}, 2023. [Online]. Available: \url{https://bit.ly/3bmE6TH}
\BIBentrySTDinterwordspacing

\bibitem{DeckerPMU}
I.~C. Decker, A.~S. e~Silva, M.~N. Agostini, F.~B. Prioste, B.~T. Mayer, and D.~Dotta, ``Experience and applications of phasor measurements to the brazilian interconnected power system,'' \emph{European Transactions on Electrical Power}, vol.~21, no.~4, pp. 1557--1573, 2011.

\bibitem{Vaz2021}
R.~Vaz, G.~R. Moraes, E.~H. Arruda, J.~C. Terceiro, A.~F. Aquino, I.~C. Decker, and D.~Issicaba, ``Event detection and classification through wavelet based method in low voltage wide-area monitoring systems,'' \emph{International Journal of Electrical Power \& Energy Systems}, vol. 130, p. 106919, 2021.

\bibitem{OperacaoSIN}
\BIBentryALTinterwordspacing
``Brazilian system operator: Operation history,'' accessed on March, 2024. [Online]. Available: \url{https://bit.ly/3VhRIqA}
\BIBentrySTDinterwordspacing

\bibitem{dado_ons_open}
\BIBentryALTinterwordspacing
``Brazilian system operator:open data,'' accessed on November, 2024. [Online]. Available: \url{http://bit.ly/4g0NzhP}
\BIBentrySTDinterwordspacing

\bibitem{curtailment_rev}
L.~Bird, D.~Lew, M.~Milligan, E.~M. Carlini, A.~Estanqueiro, D.~Flynn, E.~Gomez-Lazaro, H.~Holttinen, N.~Menemenlis, A.~Orths, P.~B. Eriksen, J.~C. Smith, L.~Soder, P.~Sorensen, A.~Altiparmakis, Y.~Yasuda, and J.~Miller, ``Wind and solar energy curtailment: A review of international experience,'' \emph{Renewable and Sustainable Energy Reviews}, vol.~65, pp. 577--586, 2016.

\bibitem{kerci2024asymmetry}
T.~Kerci and F.~Milano, ``Asymmetry of frequency distribution in power systems: Sources, impact and control,'' \emph{arXiv preprint arXiv:2405.04287.}, 2024.

\bibitem{Fanhong2020}
F.~Zeng, J.~Zhang, Y.~Zhou, and S.~Qu, ``Online identification of inertia distribution in normal operating power system,'' \emph{IEEE Transactions on Power Systems}, vol.~35, no.~4, pp. 3301--3304, 2020.

\bibitem{Misyris2023}
G.~Misyris, B.~Graham, P.~Mitra, D.~Ramasubramanian, and V.~Singhvi, ``Methodology for identifying regional inertia issues in future power grids,'' in \emph{2023 IEEE Power \& Energy Society General Meeting (PESGM)}, 2023, pp. 1--5.

\end{thebibliography}

\vspace{12pt}
\color{red}

\end{document}